\documentclass[12pt,a4paper]{article}
\usepackage{latexsym}
\usepackage{amsmath}
\usepackage[dvips]{graphicx}
\usepackage{times}
\usepackage{color}
\usepackage[colorinlistoftodos]{todonotes}

\newcommand{\bn}{\mathbf{n}}
\newcommand{\bN}{\mathbf{N}}
\newcommand{\bDe}{\mathbf{D}_{\rm e}}
\newcommand{\beps}{\boldsymbol{\varepsilon}} 
\newcommand{\ignore}[1]{}
\def\d{{\rm d}}

\usepackage{xcolor, soul}
\definecolor{orange}{rgb}{0.85,.66,0}
\sethlcolor{orange}

\title{Quasicontinuum Method\\ Extended to Irregular Lattices}
\author{Karel Mike\v{s}\footnote{\textbf{Corresponding author}: Karel Mike\v{s} \newline Address: Department of Mechanics, Faculty of Civil Engineering, Czech Technical University in Prague, Th\'{a}kurova 7, 166 29, Prague 6, Czech Republic.\newline e-mail: karel.mikes.1@fsv.cvut.cz.  
}
\and Milan Jir\'{a}sek}

\begin{document}

\maketitle
\begin{abstract}
The quasicontinuum (QC) method, originally proposed by Tadmor, Ortiz and Phillips in 1996,  
is a computational technique that can efficiently handle regular atomistic lattices by
combining continuum and atomistic approaches. In the present work, the QC method is extended 
to irregular systems of particles that represent a heterogeneous material. 
The paper introduces five QC-inspired approaches that approximate a discrete model 
consisting of particles connected by elastic links with axial interactions. Accuracy is first
checked on simple examples in two and three spatial dimensions. Computational efficiency is then
assessed by performing three-dimensional simulations of an L-shaped specimen with elastic-brittle
links. It is shown that the QC-inspired approaches substantially reduce the computational cost
and lead to macroscopic crack trajectories and global 
load-displacement curves that are very similar to those obtained by a
fully resolved particle model.
\end{abstract}





\section{Introduction}

Discrete particle models use a network of particles interacting via discrete links or connections that represent a discrete microstructure of the modeled material.
An advantage of this approach is that discrete models can naturally capture small-scale phenomena. Therefore, a variety of sophisticated discrete material models have been developed and applied
in simulations of materials such as 
paper \cite{Liu10}, textile \cite{BeexVer13,BeePee15}, fibrous materials \cite{WilBeex13,RidGon10,KulaUesa12}, woven composite fabrics \cite{PengCao04} or fiber composites \cite{Baz90}.
Extensive effort has been invested into the formulation of a discrete model of concrete \cite{Jin16,CusPel11,LilMie03,LiuDen07}.

Discrete mechanical models can accurately capture complex material response, especially localized phenomena such as damage or plastic softening.
However, they suffer by two main disadvantages.
Firstly, a large number of particles is needed to realistically describe the response of large-scale physically relevant models. This results in huge systems of equations, which are expensive to solve.
Secondly, the process of assembling of this system is also computationally expensive because all discrete connections must be individually taken into account.

Both of the aforementioned disadvantages of discrete particle models can be removed by using simplified continuous models based on one of the conventional homogenization procedures.
However, standard continuous models cannot capture localized phenomena in an objective way and
require enrichments, e.g., by nonlocal and gradient terms, which are again computationally
expensive. According to Ba\v{z}ant \cite{Baz10}, the most powerful approach to softening damage 
in the multi-scale context is a discrete (lattice-particle) simulation of the mesostructure of the entire structural region in which softening damage can occur.

Another way to reduce the computational cost of discrete particle models is a
combination of a simplified continuous model with an exact discrete description in the
parts where it is needed. Such a combination of two different approaches entails that
some hand shaking procedure is needed at the interface between the continuous and discrete
domains \cite{CurMil03}. The quasicontinuum (QC) method is a suitable technique combining the
advantages of continuous models with the exact description of discrete particle models without additional coupling procedures.

The quasicontinuum method was originally proposed by Tadmor, Ortiz and Phillips \cite{TadPhi96, TadOrt96} in 1996. The original purpose of this computational technique was a simplification of large atomistic lattice models described by long-range conservative interaction potentials. Since that time, QC methods have been widely used to investigate local phenomena of atomistic models with long-range interactions \cite{CurMil03, TadMil05}. 
Recently, the application of QC methods has been successfully extended to other lattices and interaction potentials. For example, an application of the QC method to structural lattice models of fibrous materials with short-range nearest-neighbour interactions has been developed by Beex et al.\ for conservative \cite{BeePee14} and non-conservative \cite{BeePee14a, BeePee2014c} interaction potentials including dissipation and fiber sliding as well as for planar beam lattices \cite{beex15}, still applied to regular lattices only.
An overview of applications and current directions of QC methods has been provided by Miller and Tadmor in \cite{MilTad09,MilTad02,TadMil05} and in part IV of their book \cite{TadMil11}.
In last few years, a variational formulation of dissipative QC method has been done by Roko\v{s} at al. \cite{rokos16}, a goal-oriented adaptive version of QC algorithm has been introduced in \cite{MamLar15} or a meshless QC method has been developed by Kochmann research group \cite{koch14}.
But the application of all mentioned QC methods is still restricted only to systems with regular geometry of particles.

In the present work, we extend the QC approach to irregular systems of particles with  short-range interactions by axial forces. The main idea has been tentatively presented in a conference paper \cite{MikJir15}.
Here we proceed to a more systematic evaluation of the performance of various QC formulations
applied to systems with elastic-brittle links. The proposed models are implemented 
in OOFEM \cite{oofem01,oofem12,Pat12}, an open-source object-oriented simulation platform initially
developed for finite element methods but extensible to other discretization methods.

The procedure that results from the QC method combines the following three ingredients:
\begin{enumerate}
\item
Interpolation of particle displacements is used in the regions of low interest. Only a small subset of particles is selected to characterize the behavior of the entire system. These so-called {\em repnodes} (representative nodes) are used as nodes of an underlying triangular finite element mesh, and the displacements of other particles in the region of low interest are interpolated. In the regions of high interest, all particles are selected as repnodes, in order to provide the exact resolution of the particle model. This interpolation leads to a significant reduction of the number of degrees
of freedom (DOFs) without inducing a large error in the regions of high interest.
\item
A summation rule can be applied in order to eliminate the requirement of visiting all particles during assembly of the global equilibrium equations. If such a rule is not imposed, all particles need to be visited to construct the system of equations, which makes the process computationally expensive. If the summation rule is adopted, the contribution of all particles in each interpolation triangle is estimated based on sampling of the links that surround one single particle and proper scaling of their contribution. This makes the computational process faster but some problems occur on the interface between regions of high and low interest. The piecewise linear interpolation of displacements combined with the summation rule means that the deformation is considered as constant within each interpolation element in the regions of low interest, while the deformations of individual links in the regions of high interest are evaluated exactly. Consequently, forces of nonphysical character, called the ghost forces, appear on the interface \cite{MilTad09, SheMil99}.

In our work, the summation procedure is based on homogenization of link networks contributing to the interpolation elements. Some of the links (truss elements) are selected to be processed exactly, in order to properly treat the interface between the exactly solved and interpolated domains and thus to eliminate the ghost forces.
\item
Adaptivity provides suitable changes of the regions of high interest during the simulation process. A new triangulation of the interpolation mesh could be done, but this is actually not necessary because the type of region can be changed by adding repnodes before each step. A suitable change of the regions of high interest often leads to a substantial increase of accuracy and, in several specific cases, it is necessary in order to represent the correct physical behavior, e.g., in a crack propagation process.
\end{enumerate}

\section{Methods}

\subsection{Overview}
The original QC approach was developed for regularly arranged crystal lattices, in which atoms interact at a longer distance (not just with immediate neighbors) and the interaction forces can be derived from 
suitable potentials. In regions of low interest, displacements were interpolated in a piecewise linear fashion, using a selected set of representative atoms (repatoms). In this context, imposition of an affine displacement field on the periodic crystal lattice can be interpreted as an application of the Cauchy-Born rule.  

In the present paper, we focus on discrete particle systems with short-range elastic or elastic-brittle interactions. Such systems are typically used in simulations of heterogeneous materials. Particles in these systems are distributed randomly and, in contrast to atomistic systems, do not form regular lattices, but the idea of QC can still be used.

Three approaches based on this idea are introduced here and are compared with the fully resolved particle model in two dimensions, which is considered as the reference case. Accuracy is assessed in terms of displacement and strain errors. The number and position of repnodes are adapted to achieve the optimal result.

The computational procedure consists of the following steps:
\begin{itemize}
  \item generation of particles and of connecting links,
  \item selection of repnodes and generation of interpolation elements,
  \item application of a simplification rule,
  \item assembly of global equations with repnode displacements as basic unknowns,
  \item solution of global equations (for nonlinear models using an incremental-iterative scheme),
  \item post-processing of results and error evaluation.
\end{itemize}
The details of individual steps are described in the following subsections.

\subsection{Generation of input geometry}
In the first step, the input geometry of the particle system is generated; it is specified by the position of all particles in the system and by the information which pairs of particles are connected by links. This process depends on the type of represented material.

The second step consists of repnode selection and generation of interpolation elements. There are two possible reasons why a certain particle is selected as a repnode: 
\begin{enumerate}
\item All particles located in a region of high interest are selected as repnodes to represent the ``exact''  behaviour in this region. 
\item
In regions of low interest, a sufficient number of repnodes are needed to construct the approximation of the displacements of other particles. Such repnodes represent vertices of the interpolation elements. The basic triangulation is done by the T3D mesh generator \cite{Ryp15}. Then all newly created vertices of the mesh elements are shifted to the position of the nearest particles and are labeled as repnodes.
\end{enumerate}

\subsection{Application of simplifying rule}
Once the input geometry is given, it is possible to apply a suitable simplifying rule based on the idea of QC. In this paper, five approaches using various levels of simplification are considered.

\subsubsection{Pure particle approach (A1)}
This approach does not use any simplification and corresponds to the reference model. Only the particles and links defining the particle model are used as input. Repnodes and interpolation elements are not needed. Every single particle represents a node with independent DOFs (displacements) and the links are described by 1D truss elements. Consequently, all links are taken into account explicitly and contribute directly to the internal forces and to the stiffness matrix.

This approach fully resolves the ``exact'' particle model, and the corresponding results are used as a reference solution for evaluation of accuracy and efficiency of the following simplified approaches.

\subsubsection{Hanging node approach (A2)}
\begin{figure}
\centering
    \includegraphics[trim=1.3cm 2.1cm 0cm 2.4cm, clip=true,width=0.7\textwidth]{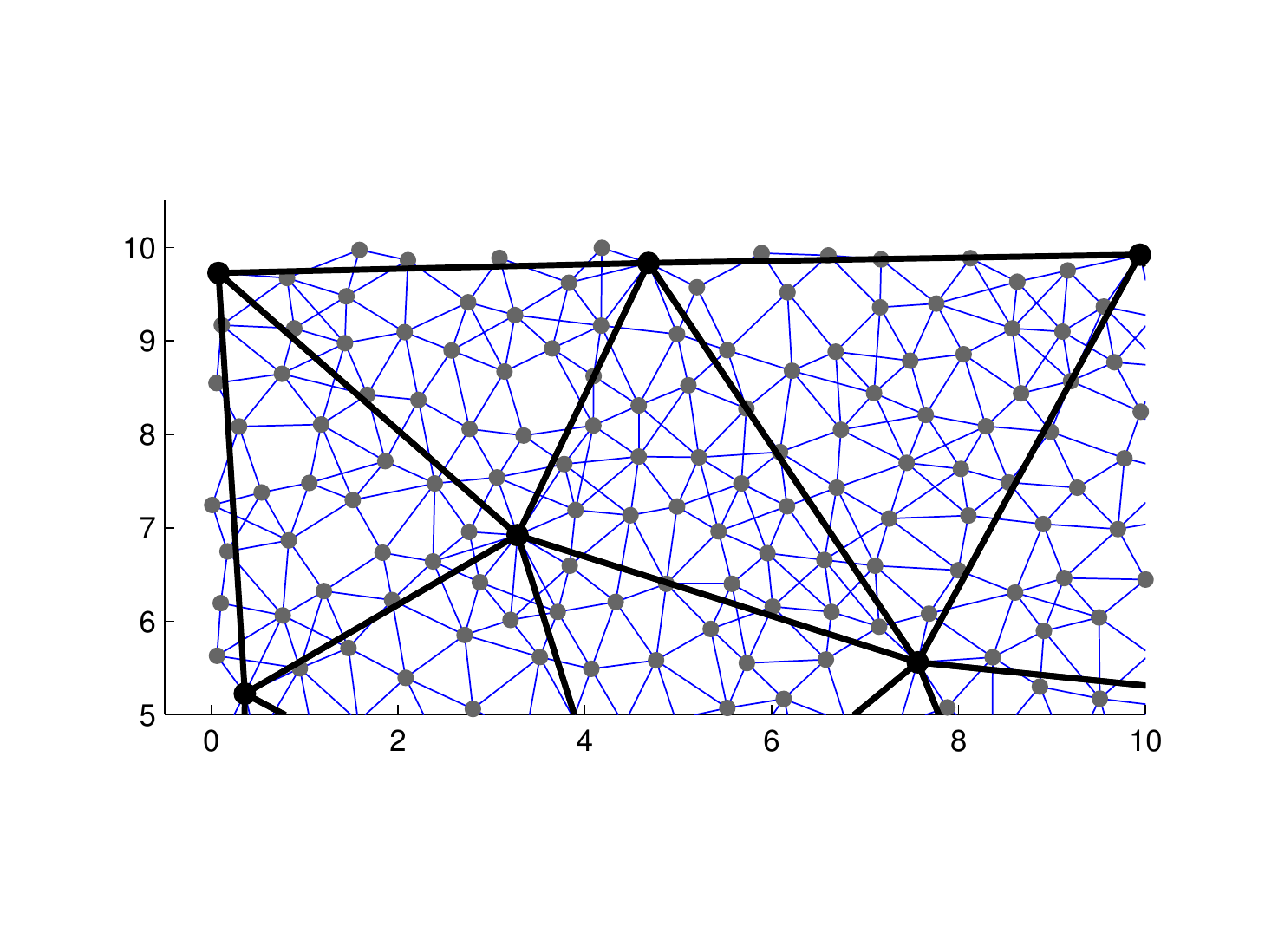}
		\caption{Example of an area of low interest, with repnodes (black circles), hanging nodes (grey circles), link connections (blue lines) and interpolation elements (triangles plotted by thick black lines)}
		\label{particles}
\end{figure}

The first technique which exploits the QC idea for simplification of the full particle model is based on approximation of DOFs of those particles that have not been selected as repnodes. Such particles are called the hanging nodes because their DOFs are not independent unknowns but are ``hanging'' on auxiliary elements with displacements interpolated from the neighboring repnodes. Triangular or tetrahedral interpolation elements with vertices at the repnodes are used here. For each hanging node, the corresponding interpolation element is found. It is either the element in which the hanging node is located, or the nearest element if the hanging node is not located in any interpolation element, which sometimes occurs at a curved part of the physical boundary of the particle system; see Figure~\ref{particles}. 

The displacement of each hanging node (grey) is a linear combination of the displacements of the vertices of the corresponding interpolation element (black). This means that DOFs of all hanging nodes are interpolated (or extrapolated) using the DOFs of repnodes as the primary unknowns. Linear interpolation is used.

All links (truss elements connecting particles) contribute to the structural stiffness matrix, but only the repnodes possess independent DOFs. The repnodes represent the nodes of an interpolation mesh, which consists of linear triangular (2D) or tetrahedral (3D) elements. These elements are used only for approximation of displacements of the nodes not selected as repnodes and do not provide a direct contribution to the internal force vector and the structural stiffness matrix.

In OOFEM implementation, the particles carrying DOFs (repnodes) are modeled as regular nodes. The particles with interpolated DOFs (hanging nodes) are represented by a special type of node for which the subset of interpolation elements can be specified. The nearest interpolation element is taken from this subset and not from all interpolation elements. This technique allows to distinguish overlapping elements on the opposite sides of a crack.

The set of links that contribute to the global equilibrium equations is the same as in the A1 approach. In regions of high interest, all nodes are repnodes and the contribution of these regions is the same as in the pure particle approach (A1).

\subsubsection{Global homogenization approach (A3)} \label{sssec:A3}
In regions of low interest,
this approach replaces the stiffness that corresponds to the links by the stiffness of 2D triangular or 3D tetrahedral elements ``filled'' by a fictitious continuum material with properties obtained by homogenization of the discrete network. Thereby, a substantial number of truss elements can be removed from the assembly procedure, and the number of operations is significantly reduced. 

In the A3 approach, only one (global) effective elastic stiffness tensor, common to all elements, is assembled from the contribution of all links.
Such a tensor can be derived from the Hill-Mandel condition \cite{Hill63,Man72},
which requires that the virtual macroscopic work density at a point  be equal to the average microscopic virtual work in a corresponding volume $V_0$ of the microstructure. This
condition can be written as
\begin{equation}
	\delta W_{mac} = \delta W_{mic} 
    \label{eq:h-m-1}
\end{equation}
where
\begin{equation}
	\delta W_{mac} =
    \boldsymbol{\sigma}:\delta\boldsymbol{\varepsilon} 
	\label{eq:Wmac}
\end{equation}
is the virtual macroscopic work density (i.e., work per unit volume)
and
\begin{equation}
	\delta W_{mic} =
    \frac{1}{V_0}\int_{V_0} \boldsymbol{\sigma}_{mic}:\delta\boldsymbol{\varepsilon}_{mic}\, \d V_0 
	\label{eq:3}
\end{equation}
is the average virtual microscopic work density in the discrete microstructure.
Here, $\boldsymbol{\sigma}$ and $\boldsymbol{\varepsilon}$ are the macroscopic
stress and strain tensors, 
and $\boldsymbol{\sigma}_{mic}$ and $\boldsymbol{\varepsilon}_{mic}$ are the microscopic
stress and strain tensors.

For a microstructure consisting of particles connected by links that transmit axial forces only,
the integral in (\ref{eq:3}) can be replaced by a sum over all links, which leads to
\begin{equation}
	\delta W_{mic} 
    = \frac{1}{V_0}\sum_{p=1}^{N_t} L_p A_p \sigma_{Np}\, \delta \varepsilon_{Np}
    = \frac{1}{V_0}\sum_{p=1}^{N_t} {F_{N}}_{p}\, \delta \Delta L_{p}
    	\label{eq:4}
\end{equation}
where $N_t$ is the number of links in volume $V_0$,
$L_p$ and $A_p$ is the length and cross-sectional area of link number $p$,
$\sigma_{Np}$ and $\varepsilon_{Np}$ is the axial stress and strain in that link,
$F_{Np} = A_p\sigma_{Np}$ is the axial force and $\Delta L_{p}=L_p \varepsilon_{Np}$
is the elongation (change of length).

In analogy to the Cauchy-Born rule used in the original quasicontinuum theory for
atomic lattices \cite{SteiEliz06}, we will use the simplifying assumption that the microscopic strains 
(actual strains as well as virtual ones) can be evaluated by projecting the macroscopic strain tensor. This assumption, in microplane theory referred to as the kinematic constraint \cite{Ozbolt01},
is written as
\begin{equation}
	\varepsilon_{Np} = \bn_p\cdot\beps\cdot\bn_p = \bN_p:\beps,
    \hskip 10mm \delta\varepsilon_{Np} = \bn_p\cdot\delta\beps\cdot\bn_p = \bN_p:\delta\beps
	\label{eq:kinematic-constraint}
\end{equation}
where $\bN_p =\bn_p \otimes \bn_p$ is a second-order tensor, introduced for convenience.
Based on (\ref{eq:Wmac})--(\ref{eq:kinematic-constraint}), the virtual work equality (\ref{eq:h-m-1}) can be rewritten as
\begin{equation}\label{eq:7}
	\boldsymbol{\sigma}:\delta\boldsymbol{\varepsilon} =  
    \frac{1}{V_0}\sum_{p=1}^{N_t} L_p A_p \sigma_{Np} \bN_p:\delta\beps
\end{equation}
Since (\ref{eq:7})
should hold for all symmetric second-order tensors $\delta\beps$, and since
$\bN_p$ is symmetric,
the macroscopic stress must be given by 
\begin{equation}\label{eq:8b}
	\boldsymbol{\sigma}=\frac{1}{V_0}\sum_{p=1}^{N_t} L_p A_p \bN_p\sigma_{Np}
\end{equation}

Formula (\ref{eq:8b})  provides a rule for the evaluation of the macroscopic stress
tensor $\boldsymbol{\sigma}$ from the microscopic stresses, in our case
from the axial stresses in individual links, $\sigma_{Np}$. The formula is generally
applicable, even to inelastic materials. In the particular case of a linear elastic material response,
the constitutive behavior of links is described by Hooke's law 
\begin{equation}
	\sigma_{Np} = E_p \varepsilon_{Np}
    	\label{eq:5}
\end{equation}
where $E_p$ is the microscopic elastic modulus of link number $p$ (often considered
as the same for all links). Substituting (\ref{eq:5}) into (\ref{eq:8b}) and exploiting
the first part of (\ref{eq:kinematic-constraint}),
we obtain
\begin{equation}\label{eq:8c}
	\boldsymbol{\sigma}=\frac{1}{V_0}\sum_{p=1}^{N_t} L_p A_p E_p \bN_p\otimes\bN_p:\beps
    =\bDe:\beps
\end{equation}
where
\begin{equation}\label{stiffness_tensor}
	\bDe=\frac{1}{V_0}\sum_{p=1}^{N_t} L_p A_p E_p \bN_p\otimes \bN_p
\end{equation}
is  the fourth-order macroscopic elastic stiffness tensor. 

In the A3 approach, the sum in (\ref{stiffness_tensor}) is taken over all links of the discrete model, and $V_0$ corresponds to the volume of the entire domain of analysis. Major and minor symmetries of the computed stiffness tensor $\bDe$ are guaranteed because all these symmetries are exhibited by the 
fourth-order tensor \mbox{$\bN_p \otimes \bN_p$}, for each $p$. Once the global stiffness tensor is evaluated, the corresponding material parameters are assigned to all 2D and 3D elements. The nature of these parameters depends on the assumed type of material (e.g., isotropic, orthotropic, or general anisotropic). For instance, if the material is supposed to be
macroscopically isotropic, the numerically evaluated stiffness tensor $\bDe$ 
is replaced by its best approximation by an isotropic stiffness tensor characterized
by two elastic constants. The details will be explained in Section~\ref{sec:homo}.

Triangular or tetrahedral elements, which were in the A2 approach considered as interpolation elements for hanging nodes, are now used directly for evaluation of the structural stiffness matrix, based on the material stiffness tensor obtained in the homogenization process. Thus, all hanging nodes with interpolated DOFs and all truss elements connecting them can be removed from the computational model. This leads to a significant reduction of the computational cost, but the elimination of links must be done carefully.

Links connecting two hanging nodes (located both in the same element or in two different elements) can be removed because their stiffness is represented by the effective stiffness of the homogenized material assigned to the elements. Links connecting one repnode and one hanging node located in the same element can be removed, too, because their stiffness is also reflected by the effective material stiffness.

\begin{figure}
\centering
\includegraphics[trim=0cm 0cm 0cm 0cm, clip=true,width=1.0\textwidth]{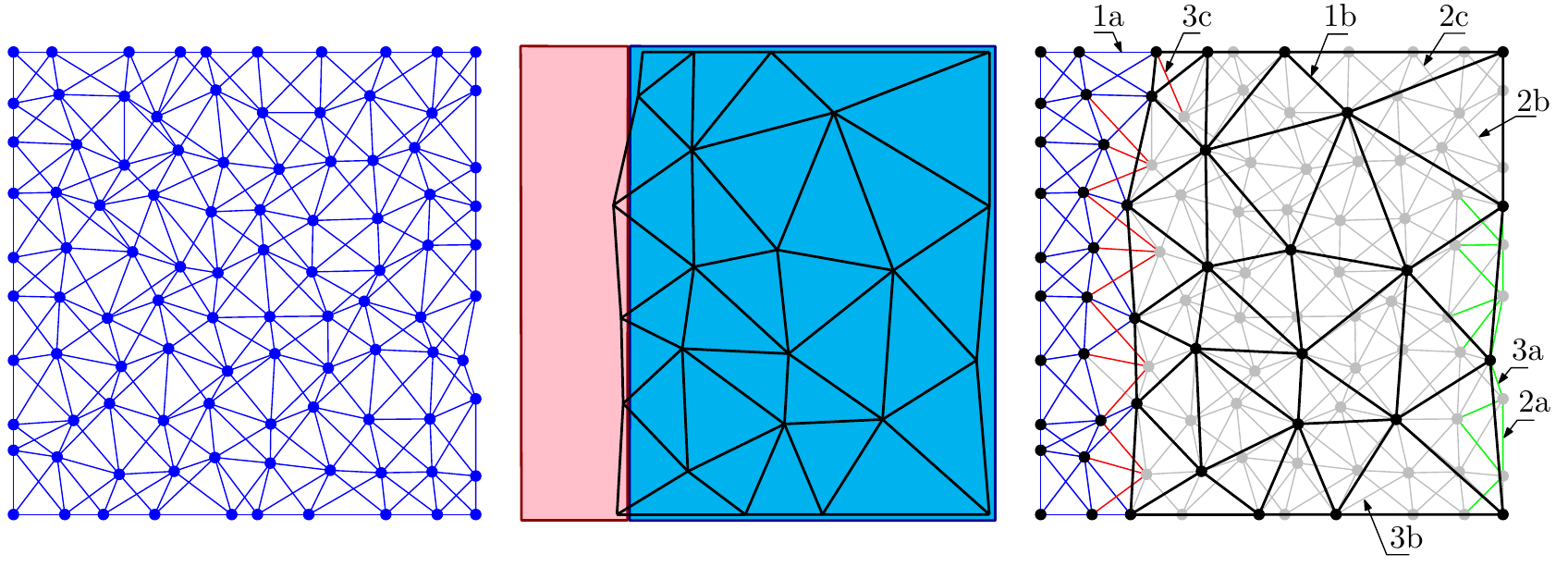}
		\caption{(a) Network of links, (b) regions of high (pink) and low (light blue) interest and interpolation
        mesh, (c) repnodes (black circles), interpolation elements (black triangles),
        links replaced by homogenized material (grey lines), and links modeled explicitly
        (blue, red and green lines) }
		\label{hanging_nodes}
\end{figure}

A special case occurs if a link passes through more than one element and connects one hanging node with one repnode. This can happen if the interpolation elements are too small, or on the interface between regions of low and high interest (see Fig.~\ref{hanging_nodes}, in which
the pink rectangle is a region of high interest and the light blue rectangle is a region
of low interest). Such links should not be removed because their stiffness is not reflected by the homogenized material. Therefore, the involved hanging nodes are kept and the contribution of the links is taken into account explicitly, in addition to the contribution of the triangular or tetrahedral elements.

The global effective elastic stiffness tensor is determined by techniques that will be
described in Section~\ref{sec:homo}.
For a general arrangement of particles and links, this tensor is considered as anisotropic.  
If the internal structure of the material exhibits no preferential directions and is expected
to be macroscopically isotropic, the numerically evaluated effective elastic stiffness is
replaced by its best isotropic approximation. To distinguish between the general anisotropic
case and the special isotropic case, the corresponding versions of the A3 approach are 
referred to as A3a and A3i.

\subsubsection{Local homogenization approaches (A4, A5)}

These approaches are refinements of A3 and aim at improving the following deficiencies:
\begin{enumerate}
  \item In the A3 approach, the effective material stiffness takes into account all links. However, certain links that partially cross homogenized elements are treated explicitly, and thus a part of their stiffness is actually accounted for twice, which artificially increases the resulting structural stiffness.
  \item The A3 approach assumes that, from the macroscopic point of view, material properties are the same at all points of the investigated domain, while in reality the local arrangement of links is variable across the domain.
\end{enumerate}

The A4/A5 approaches remove these deficiencies by identifying effective properties of the homogenized material for each element separately. 
The A4 approach treats the material as isotropic while the A5 approach accounts for local anisotropy. For computation of local material parameters, it makes sense to consider general anisotropy even if the overall material behavior is isotropic. The reason is that, for small elements, the particular local arrangement of a few links can result in a significant deviation from isotropy.

\paragraph{Evaluation of local stiffness for all homogenized elements} \mbox{}\\[2mm]
The material stiffness for each element is evaluated only from the contributions of those parts of the links that are really located inside the element. Certain links are still explicitly treated as 1D truss elements and do not contribute to the stiffness of any homogenized element.
In a loop over all links, the contribution of each link is computed according to the following set of rules, depending on the types of particles connected by the link. For better
understanding of the rules, examples of
links that correspond to individual cases (described below under labels 1ab, 2abc, 3ab) 
are provided in Fig.~\ref{hanging_nodes}c.

\begin{enumerate}
\item {\bf Repnode -- repnode}\\

\begin{enumerate}

\item
If a link is located in the region of high interest:\\
The link is taken into account explicitly as a 1D truss element.
It does not contribute to the stiffness of any homogenized element.
\item
If a link is located in the region of low interest:\\
A link connecting two repnodes in the region of low interest is always located on an edge of an interpolation element. The stiffness contribution of this link is taken into account according to Equation~\eqref{stiffness_tensor}.
The stiffness contribution is equally distributed to all elements sharing this edge.
\end{enumerate}
\item{\bf Hanging node -- hanging node}
\begin{enumerate}
\item
If one or both ends of the link are located outside of all homogenized elements:\\
The link is taken into account as a truss element.
It does not contribute to the stiffness of any homogenized element.
\item
If both ends of the link are located in the same homogenized element:\\
The stiffness contribution of this link is taken into account according to Equation~\eqref{stiffness_tensor}.
The entire contribution is assigned to the corresponding element.
\item
If the ends of the link are located in two different homogenized elements:\\
The stiffness contribution of this link is taken into account according to Equation~\eqref{stiffness_tensor}.
All elements intersected by the link are detected.
The stiffness contribution is distributed to all the detected elements in proportion to the length of the part of the link inside each element.
\end{enumerate}
\item{\bf Repnode -- hanging node}
\begin{enumerate}
\item
If the hanging node is located outside of all homogenized elements:\\
The link is taken into account as a truss element.
It does not contribute to the stiffness of any homogenized element.
\item
If both ends of the link are located in the same element:\\
The stiffness contribution of this link is taken into account according to Equation~\eqref{stiffness_tensor}.
The entire contribution is assigned to the present element.
\item
If the ends of the link are located in two different elements:\\
The link is taken into account as a truss element.
It does not contribute to the stiffness of any homogenized element.
\end{enumerate}
\end{enumerate}

If a crack is modeled by overlapping elements, the same rules can be used. It is only necessary to take a decision on which side of the crack the link is located. Based on this decision, the stiffness contribution is assigned to one of the overlapping elements.

\subsection{Homogenization}\label{sec:homo}

\subsubsection{Two-dimensional models}

Formula \eqref{stiffness_tensor} is written in tensorial notation and provides the
effective material stiffness tensor. In the actual numerical implementation, the
tensor is represented by the corresponding matrix, based on the Voigt notation.
For instance, in the two-dimensional setting, the resulting material stiffness matrix
is given by 
\begin{equation}\label{eq-stiff-mat-1}
    \mathbf{D}^{num} = 
    \left(\begin{array}{ccc} D_{1111} & D_{1122} & D_{1112} \\ D_{2211} & D_{2222} & D_{2212} \\ D_{1211} & D_{1222} & D_{1212} \end{array}\right)
\end{equation}
where $D_{ijkl}$ are components of the material stiffness tensor $\bDe$.
Note that, in two dimensions, only five of these components are independent.
Matrix $\mathbf{D}^{num}$ exhibits symmetry and, on top of that, $D_{1122}=D_{1212}$,
because formula (\ref{stiffness_tensor}) leads to
\begin{equation}
D_{1122}=\frac{1}{V_0}\sum_{p=1}^{N_t} L_p A_p E_p n_{p1}n_{p1}n_{p2}n_{p2} =
\frac{1}{V_0}\sum_{p=1}^{N_t} L_p A_p E_p n_{p1}n_{p2}n_{p1}n_{p2} =
D_{1212}
\end{equation}

If the material stiffness is considered as anisotropic, matrix $\mathbf{D}^{num}$
is used directly. This is done by the A5 approach, with local evaluation of $\mathbf{D}^{num}$
for each homogenized element separately, and also by the A3 approach, if it is decided
to use an anisotropic, globally evaluated stiffness (e.g., if the structure of the
particle model is indeed anisotropic). 

In the A4 approach, and also in the A3 approach with isotropic stiffness (referred to as A3i),
$\mathbf{D}^{num}$ is approximated by a matrix 
\begin{equation}\label{eq-stiff-mat-2}
	\mathbf{D}^{iso} = 
    \frac{E}{1-\nu^2}\left(\begin{array}{ccc} 1 & \nu & 0 \\ \nu & 1 & 0 \\ 0 & 0 & \frac{1-\nu}{2} \end{array}\right)
\end{equation}
which corresponds to the isotropic material stiffness under plane stress conditions,
with $E$ and $\nu$ denoting the Young modulus and Poisson ratio of the homogenized
material.
Alternatively, one could adopt the plane strain assumptions, which would lead to
different auxiliary values of $E$ and $\nu$ but to the same resulting matrix $\mathbf{D}^{iso}$.

Optimal values of parameters $E$ and $\nu$ are identified by minimizing
a certain measure of the difference between matrices $\mathbf{D}^{num}$ and $\mathbf{D}^{iso}$.
Several choices of such a measure are possible, but the results remain quite similar. 
The calculations presented here are based on the error measure defined as
\begin{equation}\label{eq-normA}
e(\mathbf{D}^{num},\mathbf{D}^{iso})=	
    \sum_{I=1}^{3}\left(\mathbf{v}_I^T\left( \mathbf{D}^{num} -\mathbf{D}^{iso}   \right)\mathbf{v}_I\right)^2
\end{equation}
where $\mathbf{v}_I$ is the $I$-th eigenvector of matrix $\mathbf{D}^{iso}$.
For this choice, the optimal parameters can be expressed explicitly as
\begin{eqnarray}\label{eq7}
	E&=&\frac{4(D_{1111}+2D_{1122}+D_{2222})(4D_{1111}-7D_{1122}+4D_{2222})}{33D_{1111}+6D_{1122}+33D_{2222}}\\
    \label{eq8}
	\nu&=&\frac{D_{1111}+62D_{1122}+D_{2222}}{33D_{1111}+6D_{1122}+33D_{2222}}
\end{eqnarray}
Note that the above expressions for $E$ and $\nu$ do not depend on stiffness coefficients
$D_{1112}$ and $D_{2212}$, which are always zero for isotropic materials.
If the numerically computed coefficients $D_{1112}$ and $D_{2212}$ are not small
(compared to the other coefficients), the assumption of isotropy is not appropriate
and a fully anisotropic stiffness should be used. Also note that a perfect matching
between $\mathbf{D}^{num}$ and $\mathbf{D}^{iso}$ is possible only if 
coefficients $D_{1112}$  and $D_{2212}$ vanish and the other coefficients satisfy
conditions
$D_{2222}=D_{1111}$ and $D_{1122}=D_{1111}/3$. In this case, formulae 
(\ref{eq7})--(\ref{eq8})
give $E=(8/9)D_{1111}$ and $\nu=1/3$.

An alternative error measure can be based on the standard Euclidean norm of fourth-order
tensors. The tensorial expression $(D_{ijkl}^{num}-D_{ijkl}^{iso})(D_{ijkl}^{num}-D_{ijkl}^{iso})$, with sum over $i$, $j$, $k$ and $l$ implied by the summation convention,
would be
in the Voigt notation rewritten as
\begin{equation}\label{eq-normB}
e(\mathbf{D}^{num},\mathbf{D}^{iso})=	
    \sum_{I=1}^{3}\sum_{J=1}^{3} W_{IJ}\left( D_{IJ}^{num} -D_{IJ}^{iso}   \right)^2
     = (\mathbf{D}^{num}-\mathbf{D}^{iso})\mathbf{W}(\mathbf{D}^{num}-\mathbf{D}^{iso})
\end{equation}
where $W_{IJ}$ are suitable weight coefficients 
that can be arranged into the matrix
\begin{equation}
\mathbf{W} =   \left(\begin{array}{ccc} 
    1 & 1 & 2  \\
    1 & 1 & 2  \\
    2 & 2 & 4 
    \end{array}\right)
\end{equation}
For plane stress, the corresponding expression for the optimized elastic constants is
\begin{eqnarray}\label{eq7ps}
	E&=&\frac{(D_{1111}+2D_{1122}+D_{2222})(D_{1111}+14D_{1122}+D_{2222})}{2(3D_{1111}+12D_{1122}+3D_{2222})}\\
	\nu&=&\frac{2(D_{1111}-D_{1122}+D_{2222})}{3D_{1111}+12D_{1122}+3D_{2222}}
\end{eqnarray}
For $D_{2222}=D_{1111}$ and $D_{1122}=D_{1111}/3$, we obtain again 
$E=(8/9)D_{1111}$ and $\nu=1/3$. 

Alternatively, condition $\nu=1/3$ could be imposed directly. Minimization of 
(\ref{eq-normA}) with Young's modulus considered as the only fitting variable would lead to
\begin{equation}
E=\frac{8}{81}(4D_{1111}+3D_{1122}+4D_{2222})
\end{equation}
while minimization of (\ref{eq-normB}) would give
\begin{equation}
E=\frac{2}{9}(D_{1111}+6D_{1122}+D_{2222})
\end{equation}

\subsubsection{Three-dimensional models}

In the three-dimensional setting, the resulting material stiffness matrix
is given by 
\begin{equation}\label{eq-stiff-mat-1_3d}
    \mathbf{D}^{num} = 
    \left(\begin{array}{cccccc} 
    D_{1111} & D_{1122} & D_{1133} & D_{1123} & D_{1113} & D_{1112} \\
    D_{2211} & D_{2222} & D_{2233} & D_{2223} & D_{2213} & D_{2212} \\
    D_{3311} & D_{3322} & D_{3333} & D_{3323} & D_{3313} & D_{3312} \\
    D_{2311} & D_{2322} & D_{2333} & D_{2323} & D_{2313} & D_{2312} \\
    D_{1311} & D_{1322} & D_{1333} & D_{1323} & D_{1313} & D_{1312} \\
    D_{1211} & D_{1222} & D_{1233} & D_{1223} & D_{1213} & D_{1212} 
    \end{array}\right)
    \end{equation}
Only fifteen components of the above matrix are independent because the
coefficients $D_{ijkl}$ are invariant with respect to any permutation of the subscripts.

In the A3i and A4 approaches, $\mathbf{D}^{num}$ is approximated by an isotropic stiffness matrix in the form 
\begin{equation}\label{eq-stiff-mat-2-3d}
	\mathbf{D}^{iso} = 
    \frac{E}{(1+\nu)(1-2\nu)}\left(\begin{array}{cccccc} 
    1-\nu & \nu & \nu & 0 & 0 & 0 \\
    \nu & 1-\nu & \nu & 0 & 0 & 0 \\
    \nu & \nu & 1-\nu & 0 & 0 & 0 \\
    0 & 0 & 0 & 1-2\nu & 0 & 0 \\ 
    0 & 0 & 0 & 0 & 1-2\nu & 0 \\
    0 & 0 & 0 & 0 & 0 & 1-2\nu
    \end{array}\right)
\end{equation}
The weight coefficients used in the error measure according to formula (\ref{eq-normB}) are
\begin{equation}
	\mathbf{W} = 
    \left(\begin{array}{cccccc} 
    1 & 1 & 1 & 2 & 2 & 2 \\
    1 & 1 & 1 & 2 & 2 & 2 \\
    1 & 1 & 1 & 2 & 2 & 2 \\
    2 & 2 & 2 & 4 & 4 & 4 \\
    2 & 2 & 2 & 4 & 4 & 4 \\
    2 & 2 & 2 & 4 & 4 & 4
    \end{array}\right)
\end{equation}
By minimizing this error we obtain
\begin{eqnarray}
	E&=&\frac{(a+2b)(a+11b)}{15a+39b}\\
	\nu&=&\frac{2a+b}{5a+13b}
\end{eqnarray}
where
\begin{eqnarray}
	a&=&\ D_{1111} + D_{2222} + D_{3333}  \\
	b&=&\ D_{2233} + D_{1133} + D_{1122}
\end{eqnarray}
In the special case of a perfectly isotropic matrix with $D_{3333}=D_{2222}=D_{1111}$ and $D_{2233}=D_{1133}=D_{1122}=D_{1111}/3$, the expressions can be simplified to 
$E=(5/6)D_{1111}$ and $\nu=1/4$.

\subsection{Numerical simulation}

Approaches A1--A5 described above have been implemented 
into the OOFEM open-source code \cite{oofem01,oofem12,Pat12}.
OOFEM computes displacements of repnodes and hanging nodes and strains and stresses in truss elements and homogenized planar or spatial elements. Afterwards, some post-processing procedures are required to evaluate the error of each approach and to plot the computed results.

The OOFEM input for the A4/A5 approaches is almost the same as for A3, with only one exception. An A4/A5 input file contains a large number of materials with different parameters, which are then assigned to individual elements. Numerical tests show that computations with A4/A5 are slightly slower than with A3 because approaches A4/A5 deal with more materials. However, this difference is almost negligible. The process of identifying material parameters is visibly slower for A4/A5 and the difference depends on the type of homogenization. Even if the time of the initial set-up process is taken into account,
simulations based on A3--A5  turn out to be several times faster than those based on A1 and A2.

\subsection{Error measures}
\label{subsec:Error_measures}

Accuracy of the simplifying approaches A2--A5 is evaluated by comparing the results to the 
exact approach, A1. The following error measures are used for that purpose:
\begin{enumerate}
\item
The relative stiffness error (RSE) is measured via the relative reaction error, defined as
\begin{equation}\label{errorRSE}
	RSE_{Ai} = \frac{\sum_k R_k^{(Ai)} - \sum_k R_k^{(A1)}}{\sum_k R_k^{(Ai)}}
\end{equation}
where $\sum_k R_k^{(Ai)}$ is the sum of reactions in the loading direction at all nodes $k$ with a prescribed nonzero displacement (the simulations are performed under direct displacement control).
\item
The energy error indicator (EEI) is defined as
\begin{equation}\label{eq:errorEEI}
	EEI_{Ai} = \sqrt{\frac{1}{2}\sum_{j}^{NoL} E_jA_jL_j \left(\varepsilon^{(Ai)}_j - \varepsilon^{(A1)}_j\right)^2}
\end{equation}
where $\varepsilon^{(Ai)}_j$ is the axial strain at link $j$, and the sum is taken over all links ($NoL$ denotes the number of links).
\item
The total displacement error indicator (DEI) is defined as
\begin{equation}\label{errorTDI}
	DEI_{Ai} = \sqrt{\sum_{j}^{NoN}\left\Vert \mathbf{u}^{(Ai)}_j - \mathbf{u}^{(A1)}_j\right\Vert^2}
\end{equation}
where $\mathbf{u}^{(Ai)}_j$ is the displacement vector at node $j$ and $NoN$ is the number of nodes.
\end{enumerate}

\section{Simple tests of elastic response}
\subsection{Two-dimensional periodic lattices}
\label{sec:periodic_cells}

\subsubsection{Lattice geometry and properties}

For verification of the implemented methods, the first tests are run for regular
lattices composed of periodically repeated cells. The properties of the cells are adjusted
such that the resulting macroscopic behavior be isotropic.

\begin{figure}
  \begin{center}
    \includegraphics[trim=0cm 0.0cm 0cm 0.0cm, clip=true,width=1.0\textwidth]{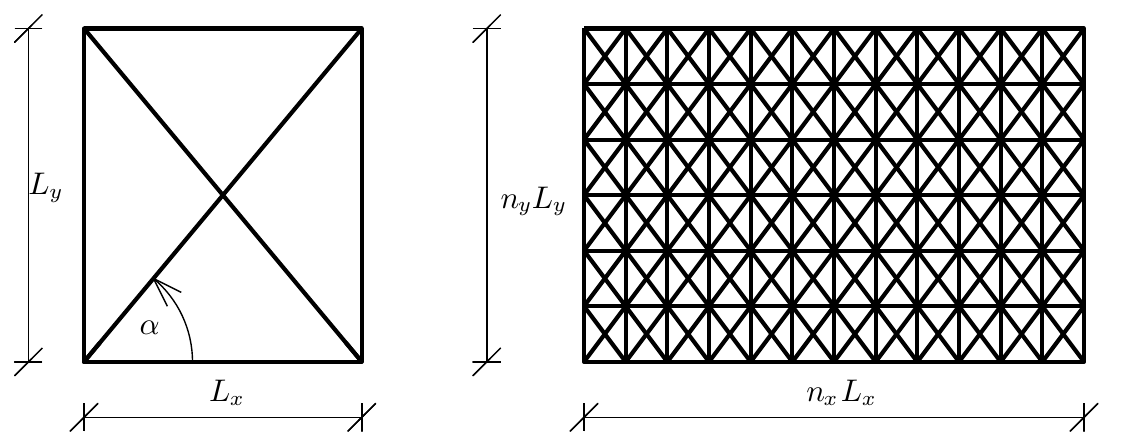}
    \caption{Geometry of characteristic cell (left) and periodic microstructure (right)}
    \label{fig-cell}
  \end{center}
\end{figure}

The microstructure is generated by periodically repeating a square-shaped basic cell 
of size $L_x\times L_y$ with
crossed diagonals; see Figure~\ref{fig-cell}. The moduli of individual links in horizontal, vertical and diagonal directions are denoted as $E_x$, $E_y$ and $E_d$, respectively. 
For simplicity, the cross-sectional area $A$ of all links within a cell 
is considered to be the same. When multiple cells are combined into a rectangular
pattern, the horizontal and vertical links on the inter-cell boundaries are merged and their resulting area
is doubled.

Due to periodicity, 
evaluation of the homogenized stiffness can be based on formula~(\ref{stiffness_tensor}) applied
to one single cell and combined with formula (\ref{eq-stiff-mat-1}).
The resulting stiffness matrix of a homogenized two-dimensional continuum is 
\begin{equation}\label{eq5}
   \mathbf{D}^{num}   =
   2\frac{A}{L_xL_yt}\left(\begin{array}{ccc} E_xL_x+E_{d}L_d\cos^4\alpha & E_{d}L_d\cos^2\alpha\sin^2\alpha & 0 \\ E_{d}L_d\cos^2\alpha\sin^2\alpha & E_yL_y+E_{d}L_d\sin^4\alpha & 0 \\ 0 & 0 & E_{d}L_d\cos^2\alpha\sin^2\alpha \end{array}\right)
\end{equation}
where $t$ is the out-of-plane thickness,
$L_d=\sqrt{L_x^2+L_y^2}$ is the length of the diagonal link,
and $\alpha=\arctan(L_y/L_x)$ is an angle characterizing the inclination of diagonals; see Fig.~\ref{fig-cell}. In general, the stiffness matrix
given by (\ref{eq5}) corresponds to an orthotropic material.

By comparing (\ref{eq5}) with formula (\ref{eq-stiff-mat-2}) for the stiffness matrix of an isotropic material under plane stress conditions, we find that the periodic cell leads to
macroscopic isotropy if the following conditions are satisfied:
\begin{eqnarray}\label{eq7x}
	E_xL_x+E_{d}L_d\cos^4\alpha &=& E_yL_y+E_{d}L_d\sin^4\alpha \\
    \label{eq8x}
	E_{d}L_d\cos^2\alpha\sin^2\alpha &=& \nu\left(E_yL_y+E_{d}L_d\sin^4\alpha\right)\\
    \nu &=& \frac{1-\nu}{2}
    \label{eq9}
\end{eqnarray}
Condition (\ref{eq9}) implies that, in the case of isotropy, the macroscopic Poisson ratio 
is restricted to $\nu=1/3$. 
From conditions (\ref{eq7x})--(\ref{eq8x})
combined with relations $L_x=L_d\cos\alpha$ and $L_y=L_d\sin\alpha$ we obtain
\begin{eqnarray}
	E_x &=& E_d\,\cos\alpha\,(3-4\cos^2\alpha) \label{eq-Ey}\\
    	E_y &=& E_d\,\sin\alpha\,(3-4\sin^2\alpha) \label{eq-Ex}
\end{eqnarray}
Finally, we can substitute $\nu=1/3$ and (\ref{eq-Ey})--(\ref{eq-Ex}) into condition
\begin{equation}
	\frac{E}{1-\nu^2} = 2\frac{A}{L_xL_yt}  ( E_xL_x+E_{d}L_d\cos^4\alpha)
\end{equation}
and link the diagonal stiffness 
\begin{equation}\label{eq13}
E_d = \frac{3tL_dE}{8A\sin 2\alpha}
\end{equation}
to geometrical parameters of the cell and to the macroscopic
elastic modulus.

For a cell of a given geometry and for a prescribed macroscopic elastic modulus, 
the characteristics of individual links in the cell can be obtained from
(\ref{eq13}) and (\ref{eq-Ey})--(\ref{eq-Ex}).
To ensure that all moduli are positive, angle $\alpha$ must be between  $\pi/6$ and
$\pi/3$, which means that the ratio $L_x:L_y$ must be between $1:\sqrt{3}$ and $\sqrt{3}:1$. 

\begin{figure}
\centering
    \includegraphics[trim=0cm 0.0cm 0cm 0.0cm, clip=true,width=1.0\textwidth]{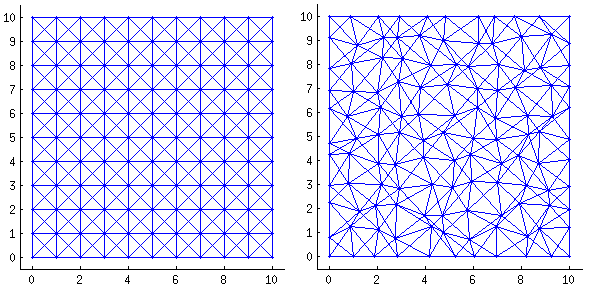}
    \caption{Examples of periodic (left) and randomized (right) microstructures with $10\times 10$ cells}
    \label{fig:patch-microstructure}
\end{figure}

\subsubsection{Direct tension test}
\begin{figure}
\centering
    \includegraphics[trim=0cm 0.0cm 0cm 0.0cm, clip=true,width=1.0\textwidth]{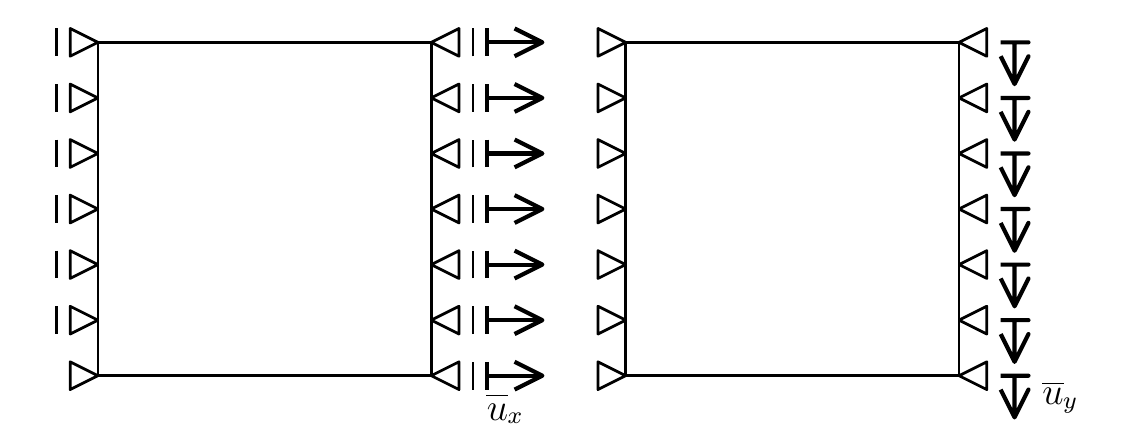}
    \caption{Boundary conditions of uniform tension test (left) and shear test (right)}
    \label{fig:patch-BC}
\end{figure}

The presented QC approaches are first subjected to simple tests in direct tension and shear
on a regular lattice shown in Fig.~\ref{fig:patch-microstructure}a,
with boundary conditions according to Fig.~\ref{fig:patch-BC}. 
The lattice is composed of $50\times 50$ periodic isotropic cells.
No region of high interest is defined in these elementary tests. 

\begin{figure}
\centering
    \includegraphics[trim=0cm 0.0cm 0cm 0.0cm, clip=true,width=1.0\textwidth]{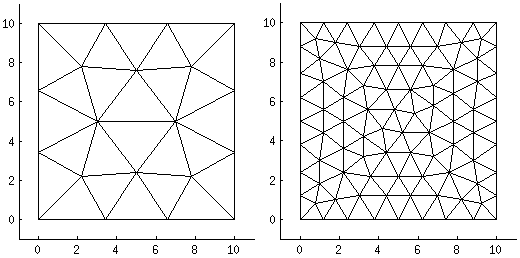}
    \caption{Examples of interpolation meshes with 3 (left) and 8 (right) elements per edge}
    \label{fig:patch-mesh}
\end{figure}

For direct tension (Fig.~\ref{fig:patch-BC}a), 
the displacements of individual particles computed using the A1 approach
(fully resolved particle model) 
exactly correspond to a linear displacement field that would arise in a homogeneous
continuum. All horizontal links are stretched in the same way and transmit the same
axial forces, and similar statements apply to the vertical links and to the diagonal links.
The purpose of this test is to check whether the simplified approaches A2--A5 lead to
the exact results. Indeed, this is the case, even on irregular meshes
shown in Fig.~\ref{fig:patch-mesh}, provided that the link stiffnesses are
tuned up such that the macroscopic behavior is isotropic. The test is
analogous to patch tests of finite elements, because it demonstrates that a solution
with a uniform strain field is captured exactly by the numerical method. 
Neither interpolation nor homogenization errors arise and all approaches pass
the patch test.

If the macroscopic behavior of the lattice is not isotropic, approaches A2 and A5 still lead to 
the exact results, while approach A3 does so only if the globally homogenized material
stiffness is considered as anisotropic. For A4, the locally homogenized material stiffness
is by definition isotropic, which induces a homogenization error. 

The fact that the global homogenization procedure gives exactly the same stiffness matrix 
$\mathbf{D}^{num}$ given by (\ref{eq5}) as the homogenization of one cell is obvious,
because the whole domain is composed of an integer number of identical cells. 
On the other hand, it is not immediately clear that the same stiffness matrix is obtained
by local homogenization over an arbitrary triangular element, including cases when edges of the element
are not aligned with lattice directions (horizontal, vertical and diagonal). 

A graphical proof of this interesting property is sketched in Fig.~\ref{fig:explanation_a},
which shows a typical triangular element and the underlying regular lattice.
The key point is that the sum of the lengths of intersections of the triangle with
 vertical links, $\sum_p L_{y,p}$, is exactly equal to the triangle area, $A_e$, divided by the horizontal
spacing between the horizontal links, $L_x$. As shown in  Fig.~\ref{fig:explanation_a}, this sum multiplied
by $L_x$ directly 
corresponds to evaluation of the triangle area by numerical integration using the
trapezoidal rule. Since the function to be integrated is piecewise linear and one of the
integration points is always located at the point where the slope changes (i.e., at the projection
of one vertex), the numerical quadrature is exact, which means that
\begin{equation}
A_e = L_x \sum_p L_{y,p}
\end{equation}
An analogous statement holds for horizontal links, for ascending diagonal 
links, and for descending diagonal links. 
Consequently, the relative
proportions of horizontal, vertical, ascending diagonal and descending diagonal links in each single triangular element are the same
as the overall relative proportions, and
formula (\ref{stiffness_tensor}) always leads to the same material stiffness
as formula (\ref{eq5}), valid for one periodic cell.
This would not be the case if the vertices
of the triangle were placed at arbitrary locations and not at grid points. 

\subsubsection{Shear test}\label{sec:shear1}

\begin{figure}
\centering
\includegraphics[trim=0cm 0.0cm 0cm 0.0cm, clip=true,width=0.7\textwidth]{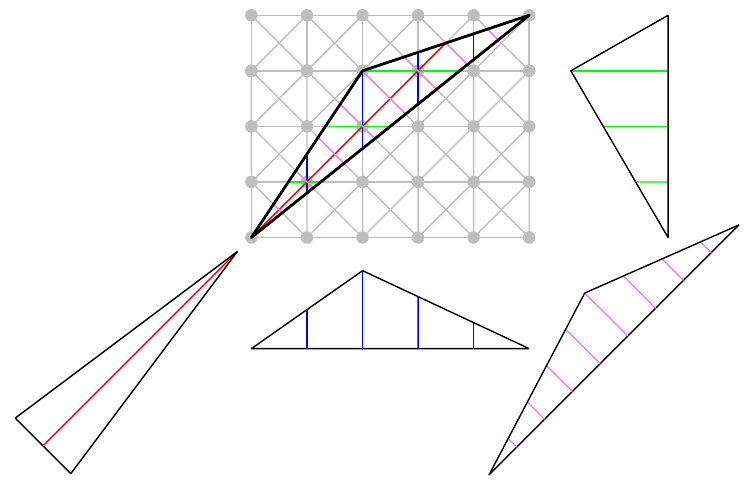}
    \caption{Graphical illustration of the reason why homogenization of a regular lattice
    over an arbitrary triangular element (with vertices at grid points) always leads to the same material stiffness}
    \label{fig:explanation_a}
\end{figure}

\begin{figure}[!ht]
  \begin{center}
    \includegraphics[trim=0cm 0.0cm 0cm 0.0cm, clip=true,width=1.0\textwidth]{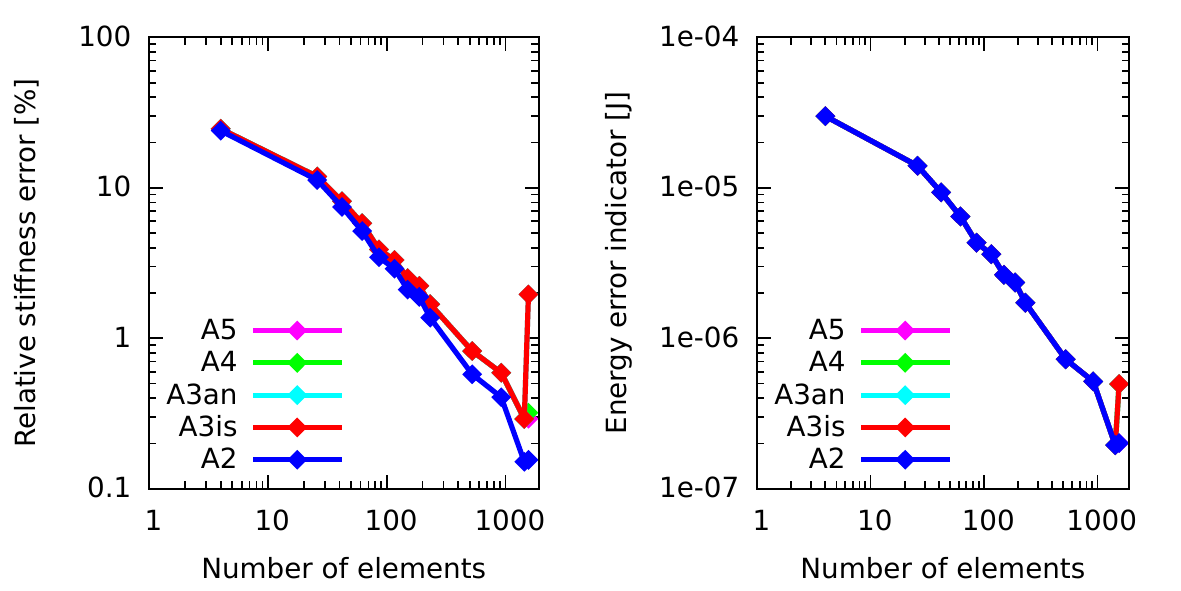}
    \caption{Relative stiffness error (a) in linear scale, (b) in logarithmic scale. 
    }
    \label{fig:cell_shear_REE+EEI}
  \end{center}
\end{figure}

For shear, the boundary conditions shown in Fig.~\ref{fig:patch-BC}b do not lead
to a uniform strain field (note that the top and bottom parts of the boundary
are considered as traction-free and their displacements are not prescribed).
The relative stiffness errors evaluated according to formula (\ref{errorRSE}) for
meshes with different numbers of elements (NoE) are 
graphically presented in Fig.~\ref{fig:cell_shear_REE+EEI}a.
The error decreases fast with increasing number of elements, which indicates that
the error is due to interpolation. The homogenization procedure, even when performed
for each triangular element separately, does not introduce any error, for the same
reasons as explained in the previous subsection on direct tension
(see Fig.~\ref{fig:explanation_a}).

\begin{figure}
\centering
\includegraphics[trim=0cm 0.0cm 0cm 0.0cm, clip=true,width=0.5\textwidth]{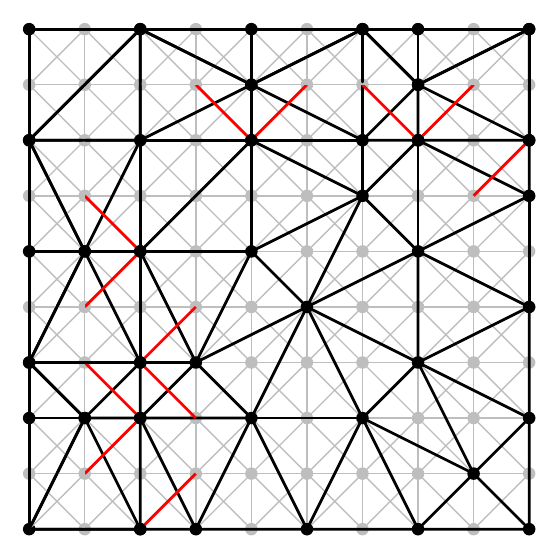}
    \caption{Example of a configuration in which some links (marked in red) are excluded from the local homogenization procedure}
    \label{fig:explanation_b}
\end{figure}

An exception is the finest interpolation mesh with 26 elements per edge,
for which the elements are not much bigger than the lattice cells.
As illustrated 
in  Fig.~\ref{fig:explanation_b}, for such fine meshes there exist
 links (marked in red) that connect a repnode with a particle located in 
 an interpolation element
not connected to that repnode. Such links are accounted for explicitly and are excluded
from homogenization, which leads to disturbances. The A5 approach with locally evaluated  anisotropic 
material stiffness is then superior to the A3 and A4 approaches, which assume isotropy.  

In terms of the stiffness error (RSE), the A2 approach (hanging nodes) gives in the present example somewhat higher accuracy
than the other approaches, even though the difference is not dramatic.
As shown in
Fig.~\ref{fig:cell_shear_REE+EEI}b, the energy error indicator (EEI)
has the same value for all approaches (A2--A5) and decreases to very low levels
as the interpolation mesh is refined.



\subsection{Two-dimensional lattices with randomized cells}

\subsubsection{Lattice geometry and properties}

In this series of tests, 
the microstructure is obtained by randomization of a periodic microstructure composed
of $50\times 50$ cells.
The randomization is achieved by random shifts of particle positions. The maximal shift values in $x$ and $y$ directions are $0.3L_x$ and $0.3L_y$, respectively. The nodes located on edges are shifted along the edges only. The final result is evaluated as an average of computations with five different randomized microstructures. An example of a randomized microstructure is plotted in Fig.~\ref{fig:patch-microstructure}b.

Patch tests have been performed for a number of triangular meshes with various sizes of elements; see Fig.~\ref{fig:patch-mesh}. 


\subsubsection{Direct tension test}

\begin{figure}
  \begin{center}
    \includegraphics[trim=0cm 0.0cm 0cm 0.0cm, clip=true,width=1.0\textwidth]    {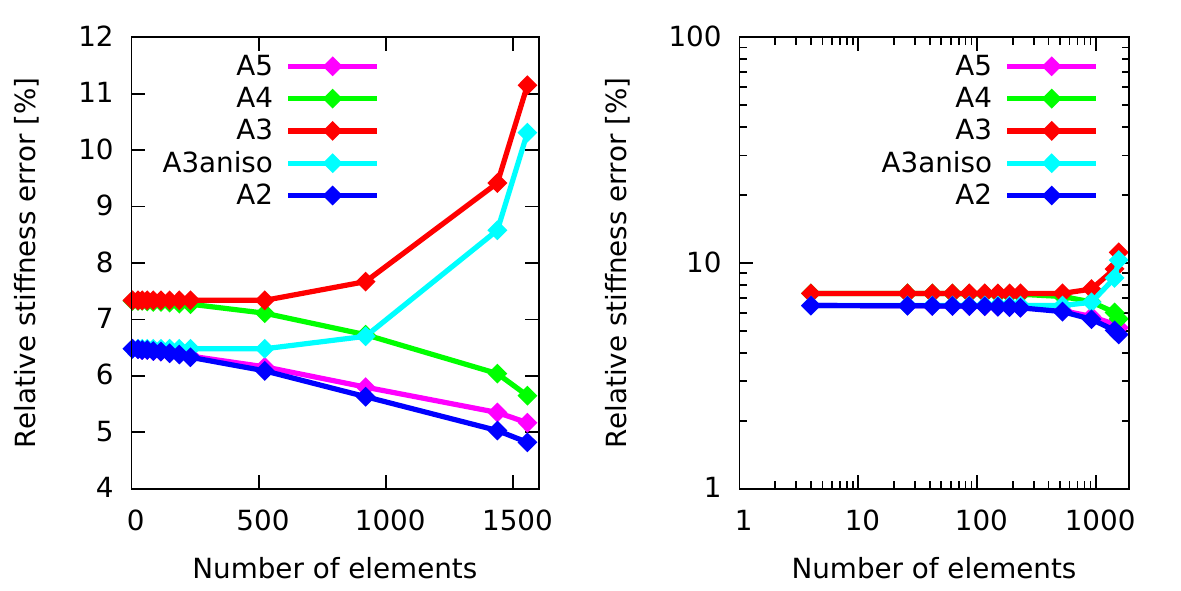}
            \caption{Relative stiffness error for a microstructure with randomized cells subjected to a uniform tension test (a) in linear scale, (b) in logarithmic scale. Isotropic parameters (for A3 and A4) obtained according to (\ref{eq-normA}).}
    
    \label{fig:tension_randomized_Stiffness_A}
  \end{center}
\end{figure}

\begin{figure}
  \begin{center}
        \includegraphics[trim=0cm 0.0cm 0cm 0.0cm, clip=true,width=1.0\textwidth]{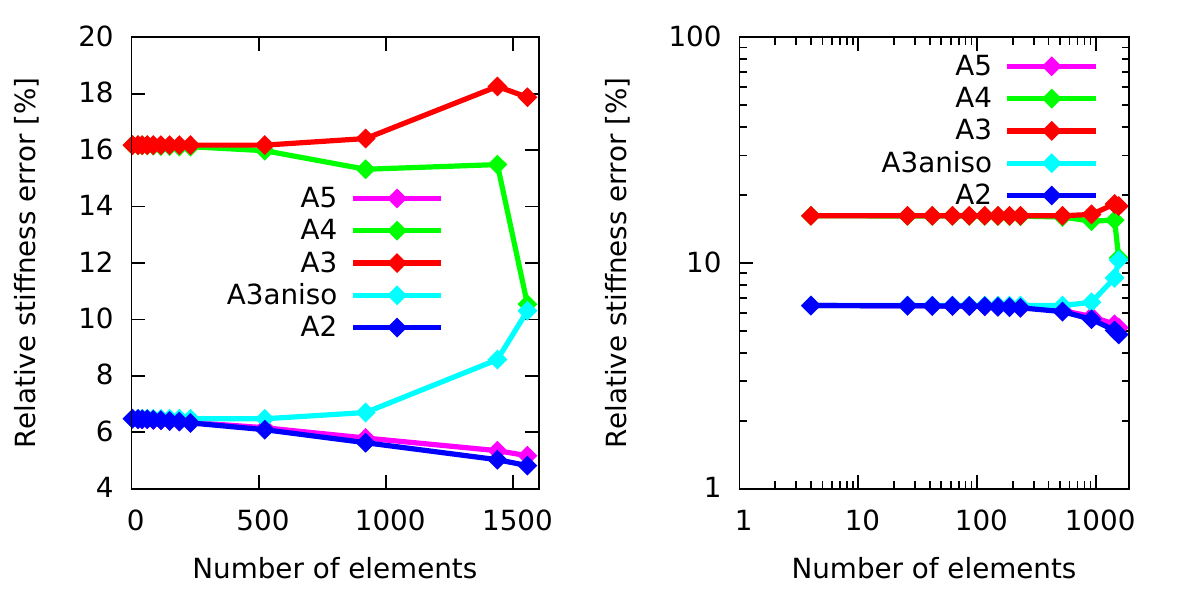}   
        \caption{Relative stiffness error for a microstructure with randomized cells subjected to a uniform tension test (a) in linear scale, (b) in logarithmic scale. Isotropic parameters (for A3 and A4) obtained according to (\ref{eq-normB}).}
    \label{fig:tension_randomized_Stiffness_B}
  \end{center}
\end{figure}

For the direct tension test, the relative stiffness errors (RSE) corresponding to 
different meshes and to approaches A2--A5 are listed in Table~\ref{tab:tension_randomized_reaction}
and graphically presented in Fig.~\ref{fig:tension_randomized_Stiffness_A}.
The energy errors (EEI) are listed in Table~\ref{tab:tension_randomized_EnergyInd}.
Isotropic parameters for A3 and A4 have been determined according to (\ref{eq-normA}).
For comparison, the RSEs are plotted again in Fig.~\ref{fig:tension_randomized_Stiffness_B}, this time with parameters A3 and A4 determined according to (\ref{eq-normB}).
The isotropic stiffness is overestimated for both matrix norms.
Nevertheless, norm (\ref{eq-normB}) provides a significantly higher error and seems to be unacceptable for application to uniform tension.  
The error induced by the A3 approach is independent of the refinement of the interpolation
mesh, except for very fine meshes. The reason is that the A3 approach deals with the same 
homogenized stiffness matrix in all elements, and if the whole sample is replaced by 
a homogeneous continuum, the prescribed boundary conditions induce
uniform strain across the sample. Such a  state of uniform strain is then captured
by all meshes because the underlying finite elements pass the standard patch test. 
Of course, the reference solution obtained using the fully resolved particle model
is somewhat different, because the microstructure is not completely regular.
The error of the A3 approach is thus nonzero; it does not depend on the mesh,
with the exception of very fine meshes, for which some links are considered
explicitly, as already discussed in Section~\ref{sec:shear1} (see Fig.~\ref{fig:explanation_b}). For approaches A4 and A5, the total error slowly decreases
as the mesh is refined. As expected, A4 is seen to be more accurate than A3, and A5
is still more accurate.

The last three columns in Table~\ref{tab:tension_randomized_reaction} present the
homogenization errors for approaches A3--A5. The homogenization error is defined as the difference between the
total error of the given approach and the total error of the A2 approach, which does
not use any homogenization. The homogenization errors are seen to be below 1\%,
and for the A5 approach to be virtually nil, with the exception of very fine meshes.
Even for coarser meshes, the homogenization error slightly increases with mesh
refinement. For approaches A3 and A4, this can be attributed to the anisotropic
character of the local arrangements of links. The deviation from isotropy is more 
pronounced on fine interpolation meshes. 


\begin{table}											
\begin{center}												
\begin{tabular}{|c|c|r|r|r|r||r|r|r|}												
\hline
\multicolumn{2}{|c|}{} & \multicolumn{4}{|c||}{Total Error} & \multicolumn{3}{|c|}{Homogenization Error} \\
\hline
NoE per edge	&	NoE	&	A2	&	A3	&	A4	&	A5	&	A3	&	A4	&	A5\\ \hline		
2	&	4	&	6.48\%	&	7.34\%	&	7.33\%	&	6.48\%	&	0.86\%	&	0.86\%	&	0.00\%		\\
3	&	26	&	6.47\%	&	7.34\%	&	7.33\%	&	6.47\%	&	0.86\%	&	0.86\%	&	0.00\%		\\	
4	&	42	&	6.46\%	&	7.34\%	&	7.33\%	&	6.46\%	&	0.87\%	&	0.87\%	&	0.00\%		\\	
5	&	62	&	6.46\%	&	7.34\%	&	7.32\%	&	6.46\%	&	0.88\%	&	0.86\%	&	0.00\%		\\	
6	&	86	&	6.44\%	&	7.34\%	&	7.31\%	&	6.44\%	&	0.89\%	&	0.87\%	&	0.00\%		\\	
7	&	116	&	6.43\%	&	7.34\%	&	7.31\%	&	6.43\%	&	0.91\%	&	0.88\%	&	0.00\%		\\	
8	&	150	&	6.40\%	&	7.34\%	&	7.30\%	&	6.41\%	&	0.94\%	&	0.90\%	&	0.01\%		\\	
9	&	188	&	6.38\%	&	7.34\%	&	7.28\%	&	6.39\%	&	0.96\%	&	0.90\%	&	0.01\%		\\	
10	&	232	&	6.33\%	&	7.34\%	&	7.27\%	&	6.35\%	&	1.00\%	&	0.94\%	&	0.02\%		\\	
15	&	524	&	6.09\%	&	7.34\%	&	7.11\%	&	6.16\%	&	1.24\%	&	1.02\%	&	0.07\%		\\	
20	&	920	&	5.63\%	&	7.67\%	&	6.73\%	&	5.80\%	&	2.04\%	&	1.10\%	&	0.18\%		\\	
25	&	1438	&	5.03\%	&	11.17\%	&	5.76\%	&	5.35\%	&	6.14\%	&	0.72\%	&	0.32\%		\\	\hline

\end{tabular}												
\caption{Relative stiffness error (RSE) for a microstructure with randomized cells subjected to a uniform tension test.}	
\label{tab:tension_randomized_reaction}     												
\end{center}												
\end{table}												

\begin{table}												
\begin{center}												
\begin{tabular}{|c|c|r|r|r|r|}												
\hline												
NoE per edge	&	NoE	&	A2	&	A3	&	A4	&	A5	\\	\hline
2	&	4	&	2.939	&	2.943	&	2.942	&	2.939	\\	
3	&	26	&	2.936	&	2.943	&	2.940	&	2.936	\\	
4	&	42	&	2.932	&	2.943	&	2.939	&	2.932	\\	
5	&	62	&	2.930	&	2.943	&	2.942	&	2.930	\\	
6	&	86	&	2.922	&	2.943	&	2.940	&	2.922	\\	
7	&	116	&	2.917	&	2.943	&	2.942	&	2.917	\\	
8	&	150	&	2.904	&	2.943	&	2.943	&	2.904	\\	
9	&	188	&	2.894	&	2.943	&	2.947	&	2.894	\\	
10	&	232	&	2.875	&	2.943	&	2.947	&	2.875	\\	
15	&	524	&	2.770	&	2.943	&	2.988	&	2.770	\\	
20	&	920	&	2.575	&	2.946	&	3.078	&	2.575	\\	\hline
\end{tabular}        												
\caption{Energy error indicator (EEI) [in $10^{-5}$] for a microstructure with randomized cells subjected to a uniform tension test.}				
\label{tab:tension_randomized_EnergyInd}     												
\end{center}          												
\end{table}												

\subsubsection{Shear test}

\begin{figure}[!ht]
  \begin{center}
    \includegraphics[trim=0cm 0.0cm 0cm 0.0cm, clip=true,width=1.0\textwidth]{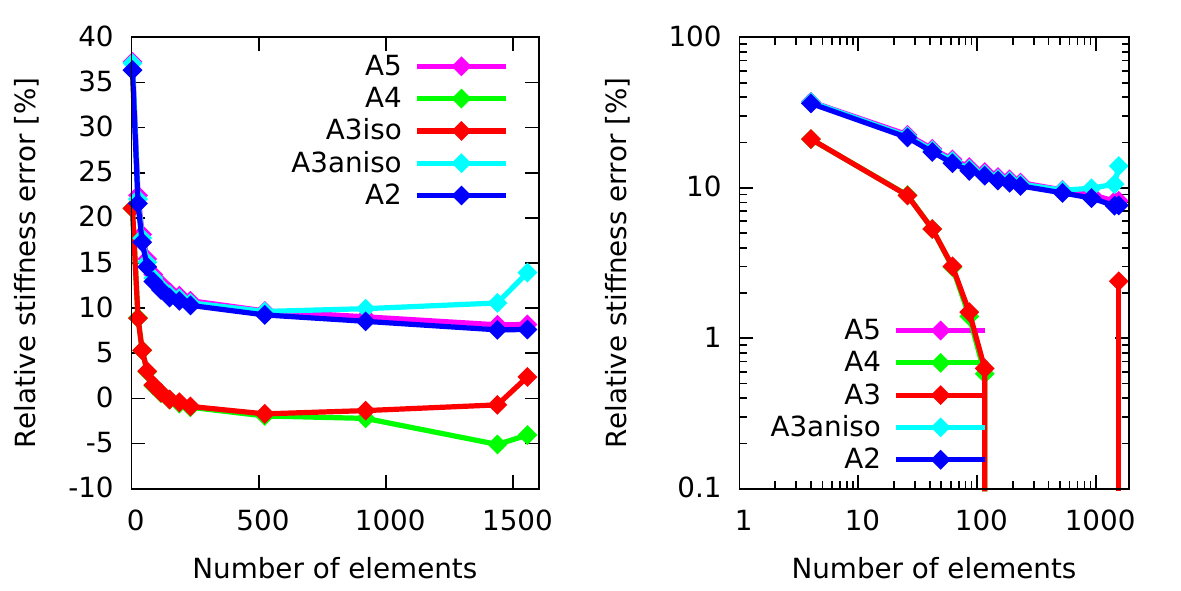}
        \caption{Relative stiffness error for a microstructure with randomized cells subjected to  shear (a) in linear scale, (b) in logarithmic scale. Isotropic parameters (for A3 and A4) obtained according to (\ref{eq-normA}).}
    \label{fig:shear_randomized_Stiffness_A}
  \end{center}
\end{figure}

\begin{figure}
  \begin{center}
    \includegraphics[trim=0cm 0.0cm 0cm 0.0cm, clip=true,width=1.0\textwidth]{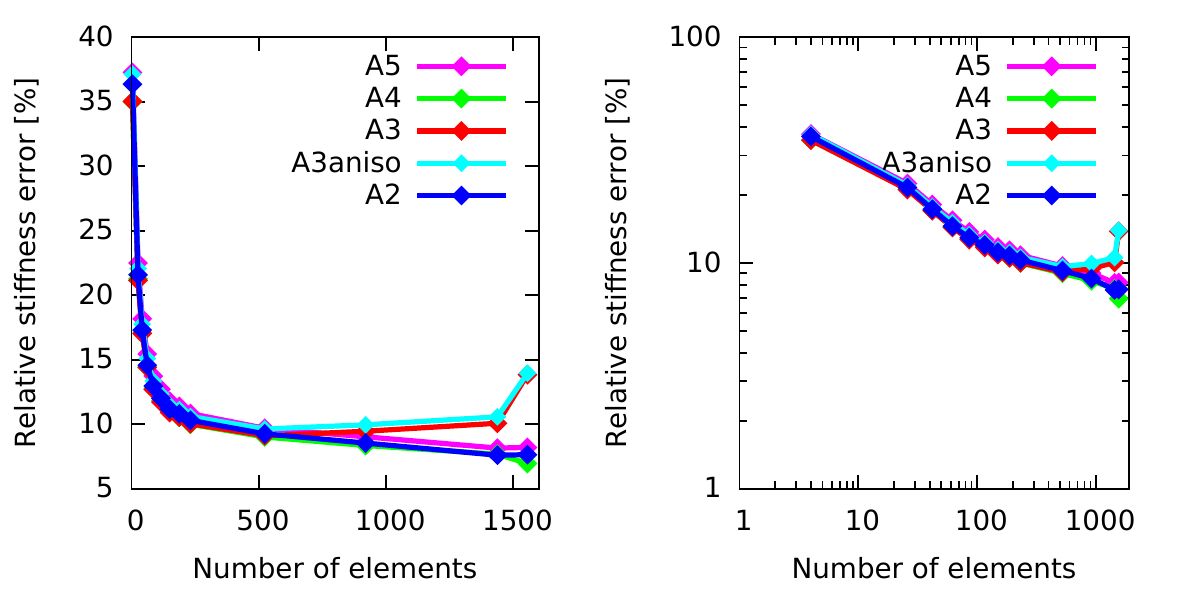}
        \caption{Relative stiffness error for a microstructure with randomized cells subjected to a shear (a) in linear scale, (b) in logarithmic scale. Isotropic parameters (for A3 and A4) obtained according to (\ref{eq-normB}).}
    \label{fig:shear_randomized_Stiffness_B}
  \end{center}
\end{figure}

For the shear test of a randomized lattice, the relative stiffness error is listed
in Table~\ref{tab:cells_reaction} and graphically presented in Fig.~\ref{fig:shear_randomized_Stiffness_A} with A3 and A4 according to the first matrix norm,
and again in Fig.~\ref{fig:shear_randomized_Stiffness_B} with A3 and A4 according to the second matrix norm.
For this shear test, the results of comparison of different matrix norms is totally opposite to the results for tension.
The second matrix norm predicts correct results comparable with other approaches,
whereas the stiffness predicted with the first norm is significantly underestimated and
the total stiffness error can even become negative, which indicates that the 
response is too compliant.
The energy error is listed in Table~\ref{tab:randomized_EnergyIdent}.
It is seen that A5 gives almost the same total error 
as A2, and the error decreases with refinement of the interpolation mesh. 
On the other hand, approaches A3 and A4 give seemingly a smaller relative stiffness error,
which can even become negative. Their energy error is larger than for A2 and A5.

Most of the error is due to interpolation. The homogenization error is smaller,
and for the A5 approach it is negligible. 

											
\begin{table}[!ht]																	
\begin{center}																	
\begin{tabular}{|c|c|r|r|r|r||r|r|r|}													
\hline																	
 \multicolumn{2}{|c|}{} & \multicolumn{4}{|c||}{Total Error} & \multicolumn{3}{|c|}{Homogenization Error} \\
\hline																	
NoE per edge	&	NoE	&	A2	&	A3	&	A4	&	A5	&	A3	&	A4	&	A5	\\	\hline
2	&	4	&	36.36\%	&	21.06\%	&	21.05\%	&	37.28\%	&	-15.30\%	&	-15.30\%	&	0.92\%	\\
3	&	26	&	21.59\%	&	8.89\%	&	8.93\%	&	22.50\%	&	-12.70\%	&	-12.66\%	&	0.90\%	\\	
4	&	42	&	17.31\%	&	5.33\%	&	5.32\%	&	18.18\%	&	-11.98\%	&	-11.99\%	&	0.87\%	\\	
5	&	62	&	14.57\%	&	3.01\%	&	2.97\%	&	15.45\%	&	-11.56\%	&	-11.61\%	&	0.88\%	\\	
6	&	86	&	12.95\%	&	1.49\%	&	1.40\%	&	13.74\%	&	-11.46\%	&	-11.55\%	&	0.78\%	\\	
7	&	116	&	12.02\%	&	0.63\%	&	0.58\%	&	12.71\%	&	-11.39\%	&	-11.44\%	&	0.70\%	\\	
8	&	150	&	11.17\%	&	-0.11\%	&	-0.14\%	&	11.78\%	&	-11.28\%	&	-11.32\%	&	0.61\%	\\	
9	&	188	&	10.80\%	&	-0.43\%	&	-0.58\%	&	11.38\%	&	-11.23\%	&	-11.38\%	&	0.58\%	\\	
10	&	232	&	10.30\%	&	-0.88\%	&	-0.98\%	&	10.83\%	&	-11.18\%	&	-11.28\%	&	0.53\%	\\	
15	&	524	&	9.25\%	&	-1.71\%	&	-1.95\%	&	9.69\%	&	-10.96\%	&	-11.20\%	&	0.44\%	\\	
20	&	920	&	8.54\%	&	-0.99\%	&	-2.03\%	&	9.03\%	&	-9.53\%	&	-10.57\%	&	0.49\%	\\	
25	&	1438	&	7.51\%	&	4.38\%	&	-0.86\%	&	8.05\%	&	-3.13\%	&	-8.37\%	&	0.54\%	\\	\hline
\end{tabular}																	
\caption{Relative stiffness error (RSE) for a microstructure with randomized cells subjected to shear.}	
\label{tab:cells_reaction}     								
\end{center}																	
\end{table}

\begin{table}												
\begin{center}												
\begin{tabular}{|c|c|r|r|r|r|}												
\hline
NoE per edge	&	total NoE	&	A2	&	A3	&	A4	&	A5	\\	\hline
2	&	4	&	4.46	&	4.52	&	4.52	&	4.46	\\	
3	&	26	&	2.65	&	2.71	&	2.72	&	2.65	\\	
4	&	42	&	2.13	&	2.19	&	2.19	&	2.13	\\	
5	&	62	&	1.79	&	1.85	&	1.86	&	1.79	\\	
6	&	86	&	1.59	&	1.65	&	1.66	&	1.59	\\	
7	&	116	&	1.48	&	1.54	&	1.55	&	1.48	\\	
8	&	150	&	1.37	&	1.44	&	1.45	&	1.37	\\	
9	&	188	&	1.33	&	1.40	&	1.41	&	1.33	\\	
10	&	232	&	1.27	&	1.34	&	1.35	&	1.27	\\	
15	&	524	&	1.14	&	1.24	&	1.28	&	1.14	\\	
20	&	920	&	1.06	&	1.22	&	1.29	&	1.06	\\	\hline
\end{tabular}        												
\caption{Energy error indicator (EEI)  [in $10^{-5}$] for a microstructure with randomized cells subjected to a shear test.}												
\label{tab:randomized_EnergyIdent}     											
\end{center}          												
\end{table}												

\clearpage{}

\begin{figure}
  \begin{center}
    \includegraphics[trim=0cm 0.0cm 0cm 0.0cm, clip=true,width=1.0\textwidth]{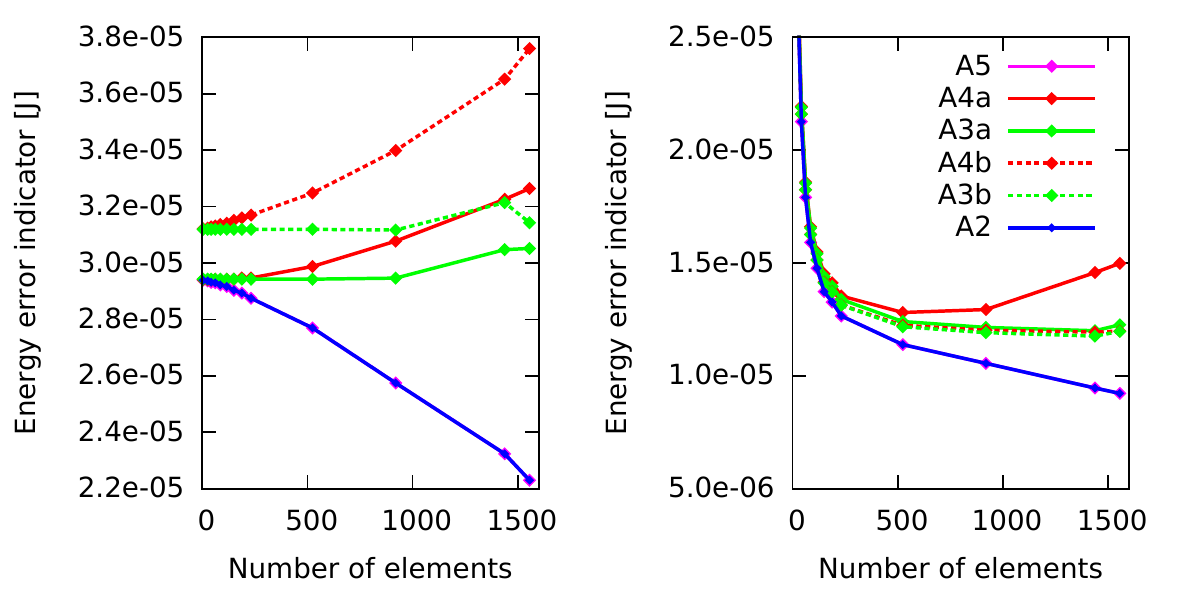}
    \caption{Energy error indicator (EEI) for a microstructure with randomized cells subjected to direct tension (left) and shear test (right)}
    \label{fig:randomizedEnergyIdent}
  \end{center}
\end{figure}

\subsection{Two-dimensional tests -- conclusions}

In partial summary, the tensile test of a regular lattice serves as a patch test that
verifies that the implementation is correct because all the approaches reproduce the
exact solution with no error. The shear test of a regular lattice shows that, upon
refinement, the error tends to zero. 

For the tensile test of a randomized lattice, the error in stiffness caused by
interpolation remains above 5\%
even on very fine meshes. For the shear test, it remains above 7.5\%. This is the
intrinsic error that needs to be accepted. Isotropic homogenization in some cases
leads to an increase of compliance, which counteracts the increase of
stiffness due to interpolation. The resulting total error in stiffness is thus in
certain cases near zero or even negative, but the energy error always remains positive.

Differences in the performance of homogenization based on error measures (\ref{eq-normA}) 
and (\ref{eq-normB}) in different patch tests indicate that
isotropic homogenization of anisotropic materials can be dangerous.
Therefore, it is better to avoid approaches A3i and A4 if the homogenized microstructure is strongly anisotropic.

\subsection{Three-dimensional patch tests}

To check the performance in three dimensions, basic patch tests are performed on cube samples composed of $24\times 24\times 24$ cells.
All presented QC approaches are again applied to direct tension and shear.
The initial periodic 3D microstructure is randomized in the same way as in 2D.
The final result is evaluated as an average of computations with five different randomized microstructures.
Parameters of isotropic stiffness (for the A3i and A4 approaches) are obtained by using matrix norm (\ref{eq-normB}) only.
Same as in the 2D case, no region of high interest is defined in these tests.

\subsubsection{Direct tension test}

\begin{figure}
  \begin{center}
    \includegraphics[trim=0cm 0.0cm 0cm 0.0cm, clip=true,width=1.0\textwidth]{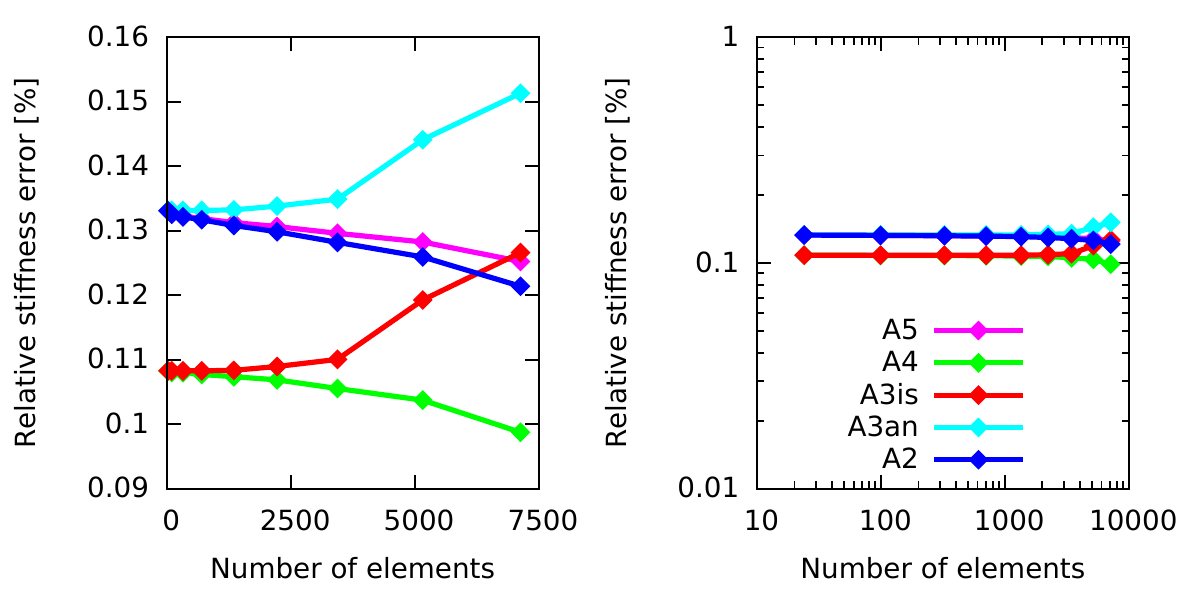}
    \caption{Relative stiffness error for a 3D randomized microstructure subjected to tension (a) in linear scale, (b) in logarithmic scale.}
    \label{fig:3dPatch-tension-RRE}
  \end{center}
\end{figure}

\begin{figure}
  \begin{center}
    \includegraphics[trim=0cm 0.0cm 0cm 0.0cm, clip=true,width=1.0\textwidth]{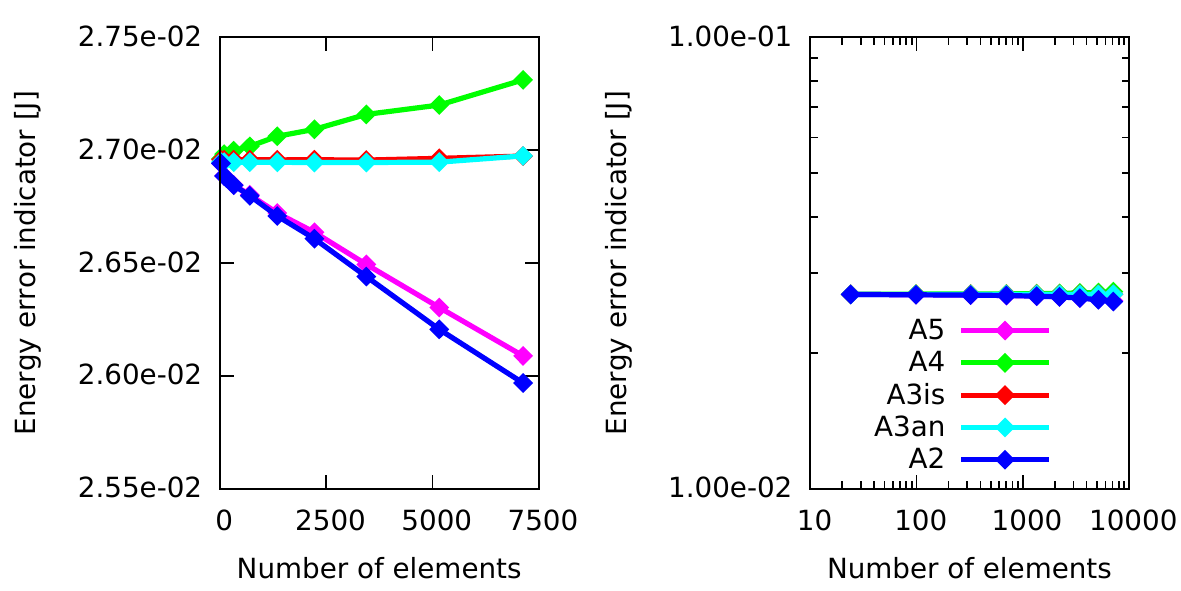}
    \caption{Energy error indicator for a 3D randomized microstructure subjected to tension (a) in linear scale, (b) in logarithmic scale.}
    \label{fig:3dPatch-tension-EEI}
  \end{center}
\end{figure}

\begin{figure}
  \begin{center}
    \includegraphics[trim=0cm 0.0cm 0cm 0.0cm, clip=true,width=1.0\textwidth]{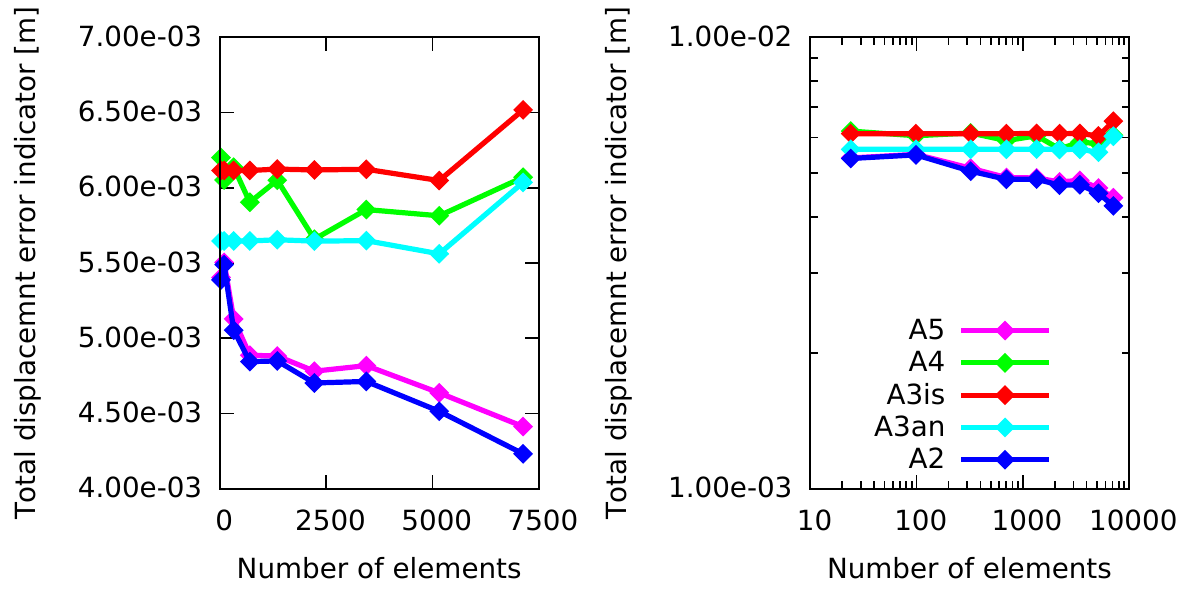}
    \caption{Total displacement error indicator for a 3D randomized microstructure subjected to tension (a) in linear scale, (b) in logarithmic scale.}
    \label{fig:3dPatch-tension-DEI}
  \end{center}
\end{figure}

For the three-dimensional direct tension test, the relative stiffness errors (RSE) corresponding to 
different meshes and to approaches A2--A5 are plotted in Fig.~\ref{fig:3dPatch-tension-RRE}.
The energy errors (EEI) are plotted in Fig.~\ref{fig:3dPatch-tension-EEI} and
the total displacement error indicator (DEI) in Fig.~\ref{fig:3dPatch-tension-DEI}.

In terms of the RSE, all approaches exhibit the same behavior as in two dimensions.
The only difference is that the homogenized isotropic stiffness is underestimated instead of overestimated.
The EEI of A3i and A3a remains constant for all mesh sizes (except the finest mesh).
This confirms the fact that, for a uniform displacement field, homogenization used by these approaches is independent of the refinement of the interpolation mesh. On the other side, the EEI of A4 increases with mesh refinement because smaller elements are statistically more anisotropic.
The order of DEI reflects the quality of homogenized procedures used in all QC approaches.

\subsubsection{Shear test}

For the three-dimensional  shear test, the RSE, EEI and DEI are plotted in Fig.~\ref{fig:3dPatch-tension-RRE},
Fig.~\ref{fig:3dPatch-tension-EEI} and Fig.~\ref{fig:3dPatch-tension-DEI}, respectively.
The performance of all approaches is quite similar. The response of A3 and A4 is too compliant due to homogenization.
Therefore these approaches appear to be more accurate than A2 and A5 in terms of RSE but not in terms of EEI and DEI.

\begin{figure}[!ht]
  \begin{center}
    \includegraphics[trim=0cm 0.0cm 0cm 0.0cm, clip=true,width=1.0\textwidth]{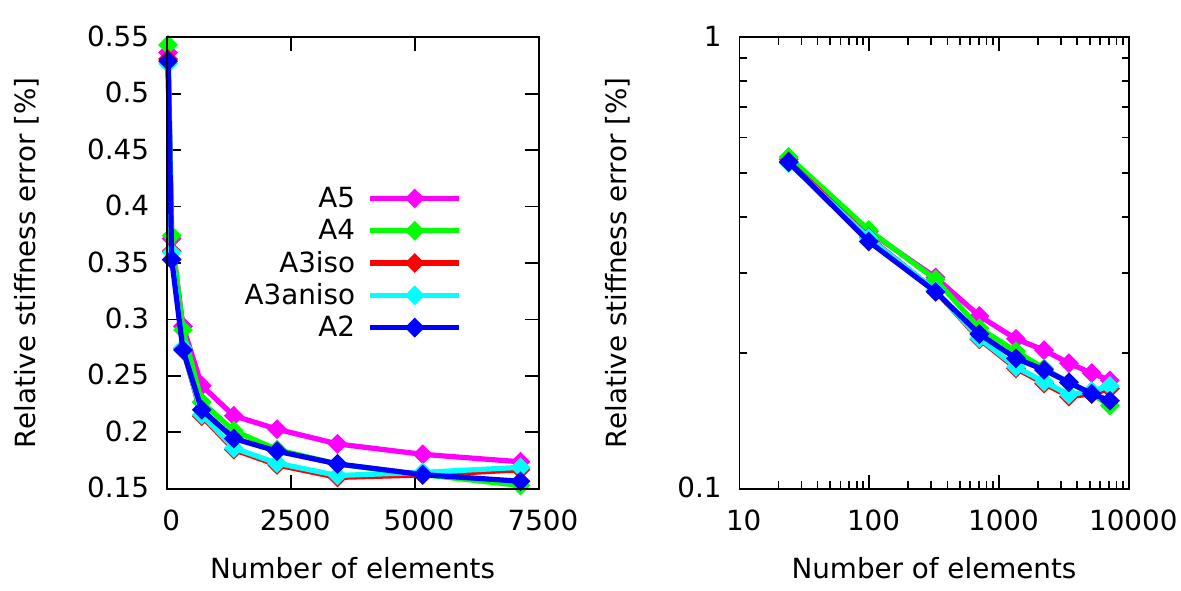}
    \caption{Relative stiffness error for a 3D randomized microstructure subjected to shear (a) in linear scale, (b) in logarithmic scale.}
    \label{fig:3dPatch-shear-RRE}
  \end{center}
\end{figure}

\begin{figure}[!ht]
  \begin{center}
    \includegraphics[trim=0cm 0.0cm 0cm 0.0cm, clip=true,width=1.0\textwidth]{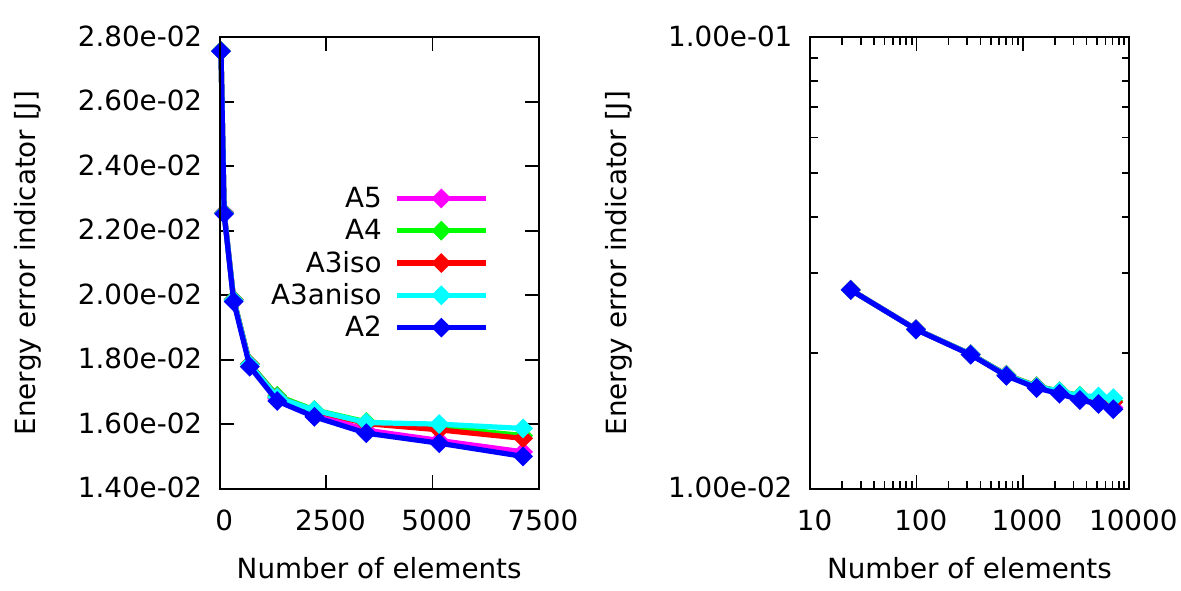}
    \caption{Energy error indicator for a 3D randomized microstructure subjected to shear (a) in linear scale, (b) in logarithmic scale.}
    \label{fig:3dPatch-shear-EEI}
  \end{center}
\end{figure}

\begin{figure}[!ht]
  \begin{center}
    \includegraphics[trim=0cm 0.0cm 0cm 0.0cm, clip=true,width=1.0\textwidth]{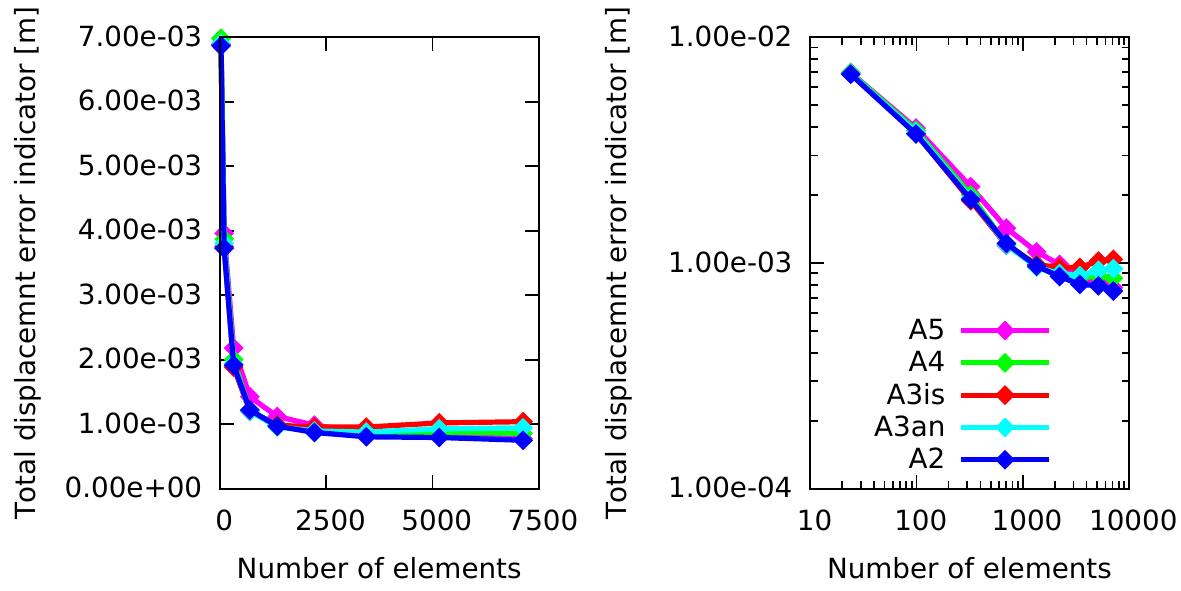}
    \caption{Total displacement error indicator for a 3D randomized microstructure subjected to shear (a) in linear scale, (b) in logarithmic scale.}
    \label{fig:3dPatch-shear-DEI}
  \end{center}
\end{figure}

\subsubsection{Three-dimensional tests -- conclusions}
Three-dimensional patch tests have shown that local anisotropic homogenization (i.e., the A5 approach) provides results with an almost zero homogenization error, but an intrinsic stiffness error around 15\% due to interpolation must be taken into account.


\section{Failure simulations}


To assess the efficiency and accuracy of QC-based approaches in applications that involve 
inelastic material response, failure of an L-shaped specimen is simulated in three dimensions.  
The behavior of links that connect particles is assumed to be elastic-perfectly brittle, with link breakage occurring at a critical level of tensile strain.

The L-shaped specimen shown in Fig.~\ref{fig:3dCracking-geometry2} has
dimensions $300\times 300\times 100$ mm. It is fixed at the bottom section and loaded by
prescribed vertical and horizontal displacements imposed at the left end section. 
As a result, the non-convex corner is opened, the upper part of the specimen is bent
and the horizontal part is twisted.

The microstructure is generated with a density of 20 nodes along the short edge and randomized. This procedure results in 38,400 particles (with 113,200 unknown DOFs) connected by 324,672 links.
Material parameters are considered to be the same for all links.

In the simulation, the prescribed displacement is increased until the critical value of tensile strain is reached in the most stretched link. Subsequently, the link is removed and the loading is imposed again. Repeating this process results in a series of link breakages that define the macroscopic crack trajectory.

\subsection{Cylindrical fully resolved domain}

\begin{figure}
  \begin{center}
    \includegraphics[trim=0cm 0.0cm 0cm 0.0cm, clip=true,width=1.0\textwidth]{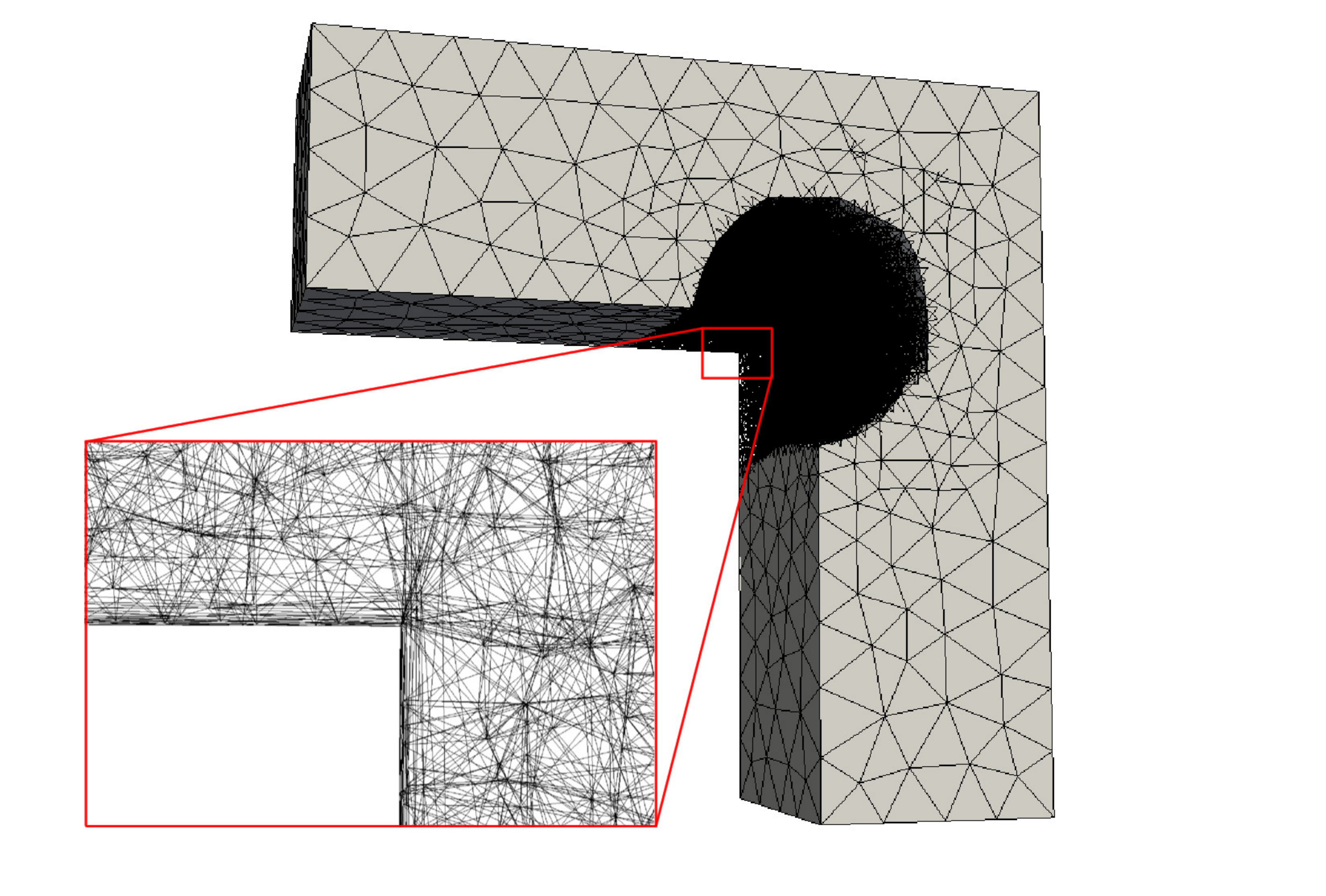}
    \caption{Geometry of interpolation mesh and the fully resolved domain with a detail of microstructure.}
    \label{fig:3dCracking-geometry2}
  \end{center}
\end{figure}

For the purpose of this test, a cylindrical fully resolved domain (FSD) is specified around the non-convex corner; see Fig.~\ref{fig:3dCracking-geometry2}. The FSD radius is set to 50 mm, i.e., to one half of the length of the shortest specimen edge. This FSD occupies 11.8 \% of the volume of the entire domain. Basic characteristics of models used by simplified QC approaches are listed in Table~\ref{tab:3d-L-beam-prop-1000}.

\begin{figure}
  \begin{center}
    \includegraphics[trim=0cm 0.0cm 0.5cm 0.0cm, clip=true,width=0.49\textwidth]{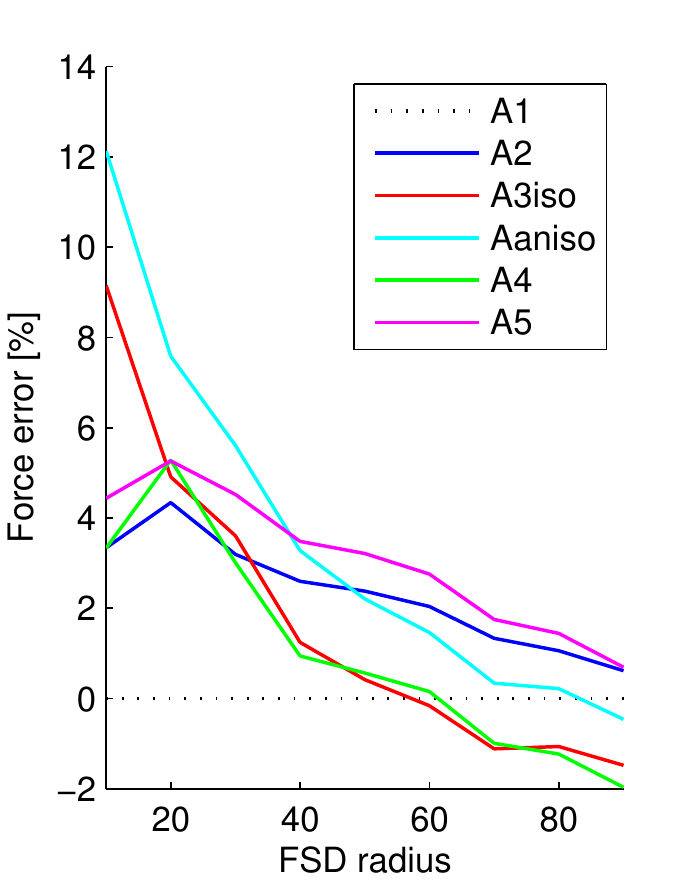}
        \includegraphics[trim=0cm 0.0cm 0.5cm 0.0cm, clip=true,width=0.49\textwidth]{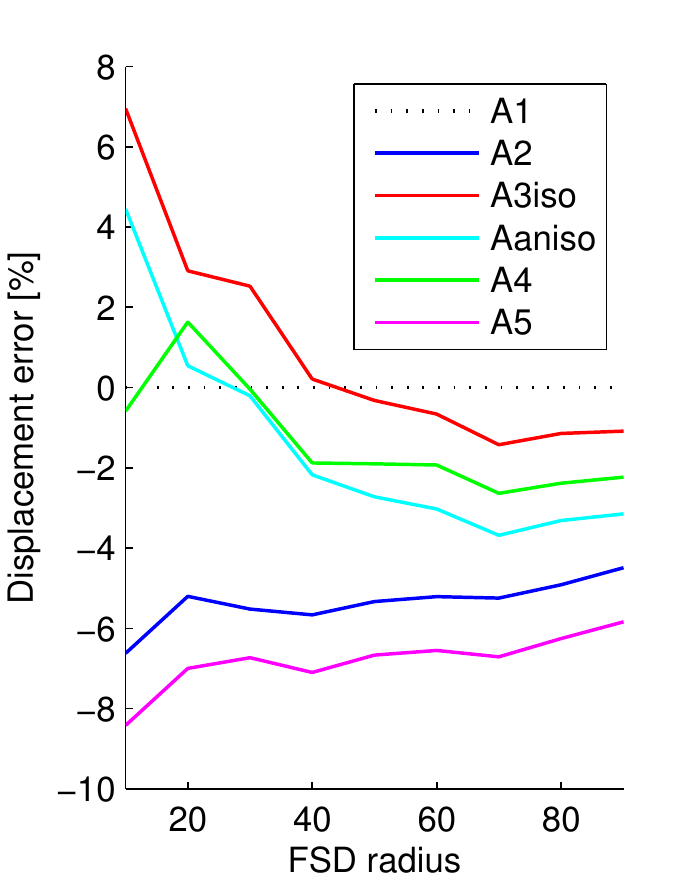}
    \caption{Relative force error (left) and relative prescribed displacement error (right) for different radii of the fully resolved domain.}
    \label{fig:L-beam-FSD-error-F+u}
  \end{center}
\end{figure}

\begin{figure}
  \begin{center}
    \includegraphics[trim=0cm 0.0cm 0.5cm 0.0cm, clip=true,width=0.49\textwidth]{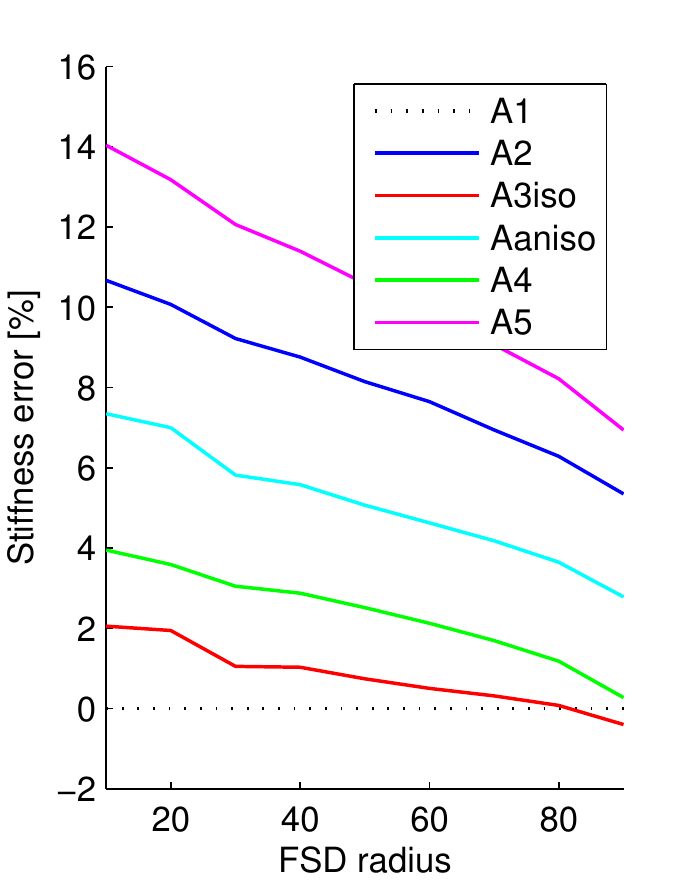}
        \includegraphics[trim=0cm 0.0cm 0.5cm 0.0cm, clip=true,width=0.49\textwidth]{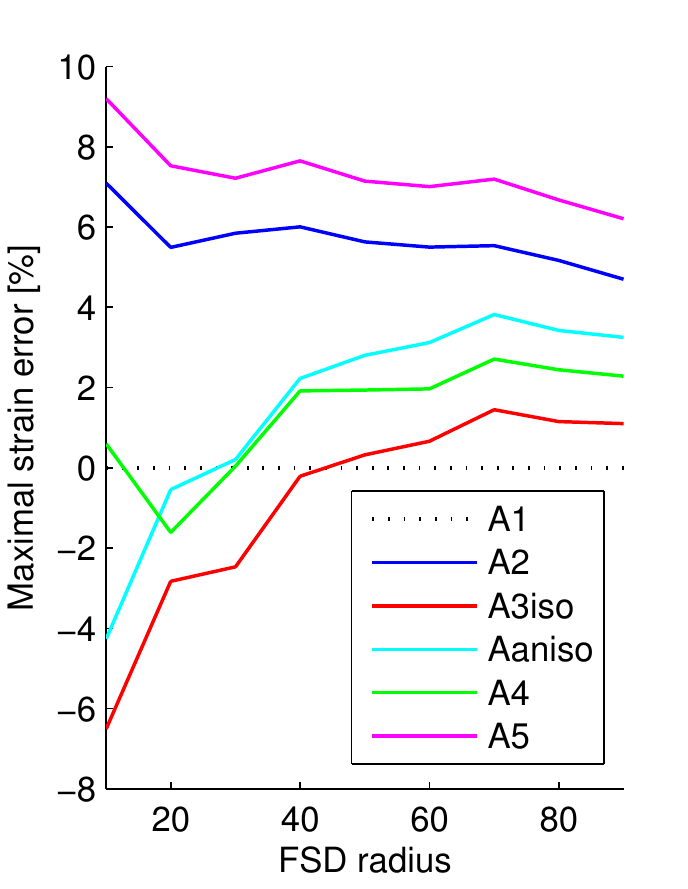}
    \caption{Relative stiffness error (left) and relative error of strain in the most loaded link (right) for different radii of the fully resolved domain.}
    \label{fig:L-beam-FSD-error-k+eps}
  \end{center}
\end{figure}

\begin{figure}
  \begin{center}
    \includegraphics[trim=0cm 0.0cm 0.5cm 0.0cm, clip=true,width=0.49\textwidth]{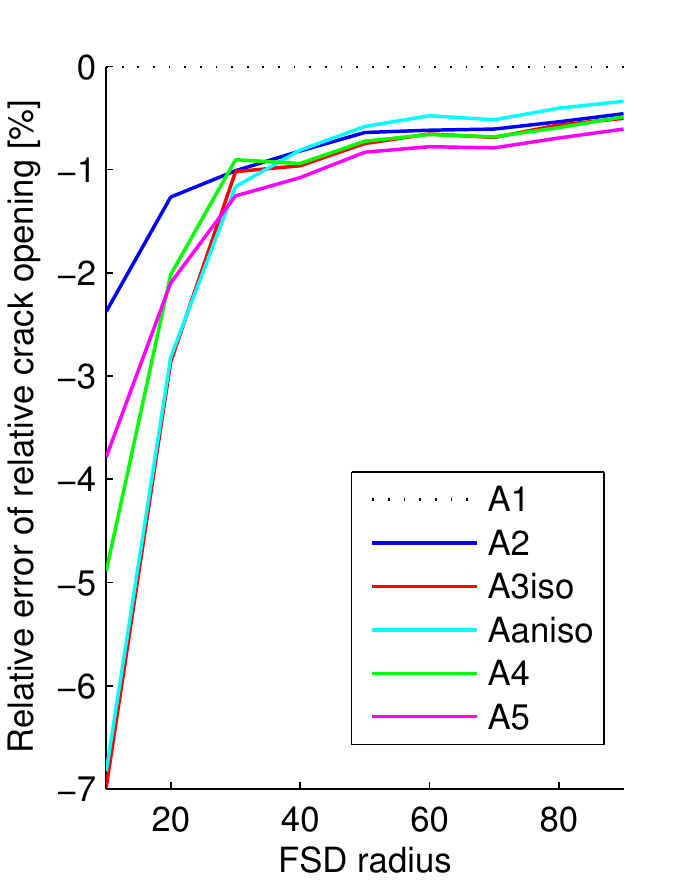}
        \includegraphics[trim=0cm 0.0cm 0.5cm 0.0cm, clip=true,width=0.49\textwidth]{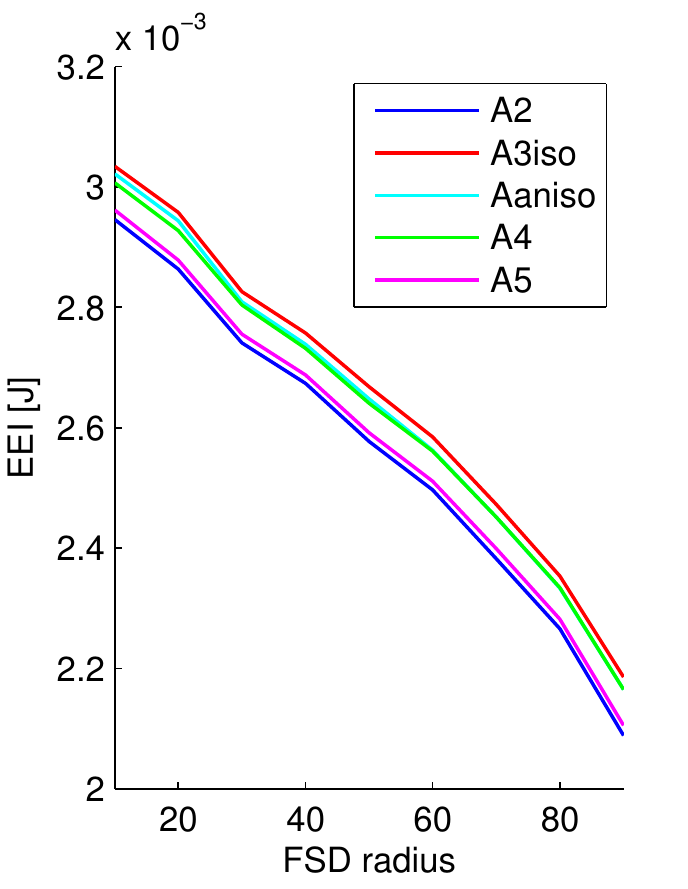}
    \caption{Crack opening error at the moment when critical strain is reached (left) and energy error indicator (right) for different radii of the fully resolved domain.}
    \label{fig:L-beam-FSD-error-op+en}
  \end{center}
\end{figure}

The simulation proceeds until 1000 links are cracked. Relative errors evaluated by comparing the results of simplified approaches to the full particle model (approach A1) are depicted in Fig.~\ref{fig:L-beam-FSD-error-F+u} and \ref{fig:L-beam-FSD-error-k+eps}. Crack opening error and the energy error indicator (EEI) defined in (\ref{eq:errorEEI}) are depicted in Fig.~\ref{fig:L-beam-FSD-error-op+en}.
The ranking  of individual QC approaches according to EEI in the first step reflects the quality of homogenization methods used by these approaches. 
On the other side, in the last step, the ranking of approaches is different and depends on which links are cracked.

The capability of all simplified approaches A2-A5 to correctly predict the link that will break next
(i.e., the link with maximum strain) depends on the current microstructure. The numbers of incorrectly cracked links during the failure process are compared in Table~\ref{tab:3d-L-beam-cracking}. The numbers in the line denoted as $500/1000$ indicate how many of the first 500 cracked links for the given approach are not found in the list of the first 1000 cracked links in the exact solution (approach A1). For simplified approaches, the precise sequence of cracked links is not always the same but just a small number of cracked links are predicted incorrectly.
Even though a few incorrectly cracked links appear, the simplified approaches are able to predict a correct overall crack trajectory, provided that the FSD is selected properly.

Another important quantity is 
the maximum value of loading force $F_{\rm max}$ observed in the force-displacement diagram. The results are shown in Table~\ref{tab:3d-L-beam-maxF}. For all simplified approaches, the maximum force is predicted in the 35th step, while the full particle model (approach A1) gives the maximum force in the 36th step. However, relative errors in $F_{\rm max}$ as well as relative errors in the corresponding displacement are just a few percent.

Computational times consumed by individual components of various QC-based approaches are summarized in Table~\ref{tab:3d-L-beam-time-simplification}. Two most demanding procedures are the assignment of interpolation elements to all particles and the assembly of individual stiffness tensors from all elements.
Searching for the interpolation element is done independently for each particle.
The distribution of link stiffnesses to individual tensors is also done independently for each link.
Therefore, parallelization of the loop over particles or links can easily be envisaged with an almost ideal expected speed-up.

The computational times of one step and of the whole simulation are shown in Table~\ref{tab:3d-L-beam-time-1000}. The QC approaches are able to reduce the computational time of one step more than ten times.
Even if the computational time needed for the initial simplification is high, the total simulation time is significantly reduced if the total number of steps is huge.

\begin{table}						
\begin{center}							
\begin{tabular}{|l|r|r|r|}							
\hline							
~ 		&	A1		&	A2		&	A3-A5	\\ \hline
NoP    	&	38,400	&	38,400	&	6,752	\\
NoL    	&	324,672	&	324,672	&	39,695	\\
NoE    	&	0		&	4,457	&	4,457	\\
NoRN   	&	38,400	&	5,388	&	5,388	\\
NoHN   	&	0		&	33,012	&	1,364	\\
DOFs   	&	113,200	&	16,014	&	16,014	\\ \hline

\end{tabular}							
\caption{Numbers of particles, links, elements, repnodes, hanging nodes and DOFs for various QC approaches.}							
\label{tab:3d-L-beam-prop-1000}							
\end{center}							
\end{table}


\begin{table}								
\begin{center}									
\begin{tabular}{|l|c|c|c|c|c|}									
\hline									
~								&	A2		&	A3i		&  A3a	&	A4 	& A5	\\ \hline
Incorrectly outside FSD			&	4	&	6	&	13	&	10	&	7	\\
Incorrectly cracked 1000/1000	&	48	&	60	&	55	&	86	&	52	\\
Incorrectly cracked 500/1000	&	15	&	4	&	4	&	24	&	4	\\ \hline
\end{tabular}									
\caption{Numbers of incorrectly cracked links.}				
\label{tab:3d-L-beam-cracking}									
\end{center}									
\end{table}	

\begin{table}						
\begin{center}							
\begin{tabular}{|l|r|r|r|}							
\hline							
~	&	$F_{\rm max}$ error	&	${F_{\rm max}}$ displacement error	&	$F_{\rm max}$ step number	\\ \hline
A1  	&	-		&	-		&	36	\\
A2  	&	3.2\%	&	-4.5\%	&	35	\\
A3i  	&	1.1\%	&	0.4\%	&	35	\\
A3a 	&	2.9\%	&	-2.1\%	&	35	\\
A4  	&	1.4\%	&	-1.1\%	&	35	\\
A5  	&	4.3\%	&	-5.7\%	&	35	\\ \hline
\end{tabular}							
\caption{Tensile strength errors.}							
\label{tab:3d-L-beam-maxF}							
\end{center}							
\end{table}							

\begin{table}								
\begin{center}									
\begin{tabular}{|l|c|c|c|c|}									
\hline									
~								&	A1	&	A2		&	A3i, A3a	&	A4, A5	\\ \hline
Generate interpolation mesh		&	-	&	0.09s	&	0.09s	&	0.09s	\\
Transform mesh to particles		&	-	&	0.19s	&	0.19s	&	0.19s	\\
Find element for all particles	&	-	&	0.92s	&	0.92s	&	0.92s	\\
Global stiffness tensor			&	-	&	-		&	0.15s	&	-		\\
Connectivity table				&	-	&	-		&	-		&	0.17s	\\
Individual stiffness tensors	&	-	&	-		&	-		&	19.27s	\\ \hline
Total simplification time		& 0.0s	&	1.20s	&	1.35s	&	20.64s	\\ \hline
\end{tabular}									
\caption{Time consumption of QC simplifications (i.e., of the preparatory stage before the actual simulation).}		
\label{tab:3d-L-beam-time-simplification}	
\end{center}									
\end{table}

\begin{table}								
\begin{center}								
\begin{tabular}{|c|c|c|c|} \hline
~	&	{Time of QC}	& Computational time of one step 	& Total comp. time\\
~	&	simplification	& step time / solver time			& of 1000 steps	 \\ \hline
A1	&	-		&	8.2s / 5.72s (1.00 / 1.00)	&	2h 16m 40s (1.00) \\
A2	&	1.2s	&	13.66s / 0.47s (1.67 / 0.08)&	3h 47m 41s (1.67)\\
A3i	&	1.35s	&	0.85s / 0.38s	(0.10 / 0.07)&	14m 11s (0.10)\\
A3a	&	1.35s	&	0.85s / 0.39s	(0.10 / 0.07)&	14m 11s (0.10)\\
A4	&  20.64s	&	0.85s / 0.39s	(0.10 / 0.07)&	14m 30s (0.11)\\
A5	&  20.64s	&	0.85s / 0.38s	(0.10 / 0.07)&	14m 30 (0.11)\\ \hline
\end{tabular}								
\caption{Computational times consumed by various QC approaches for a conjugate gradient (CG) solver with incomplete Cholesky preconditioning and symmetric compressed column matrix storage scheme.}
\label{tab:3d-L-beam-time-1000}								
\end{center}								
\end{table}

\subsection{Fully resolved domain for crack propagation}
\begin{figure}
  \begin{center}
    \includegraphics[trim=0cm 0.0cm 0cm 0.0cm, clip=true,width=0.50\textwidth]{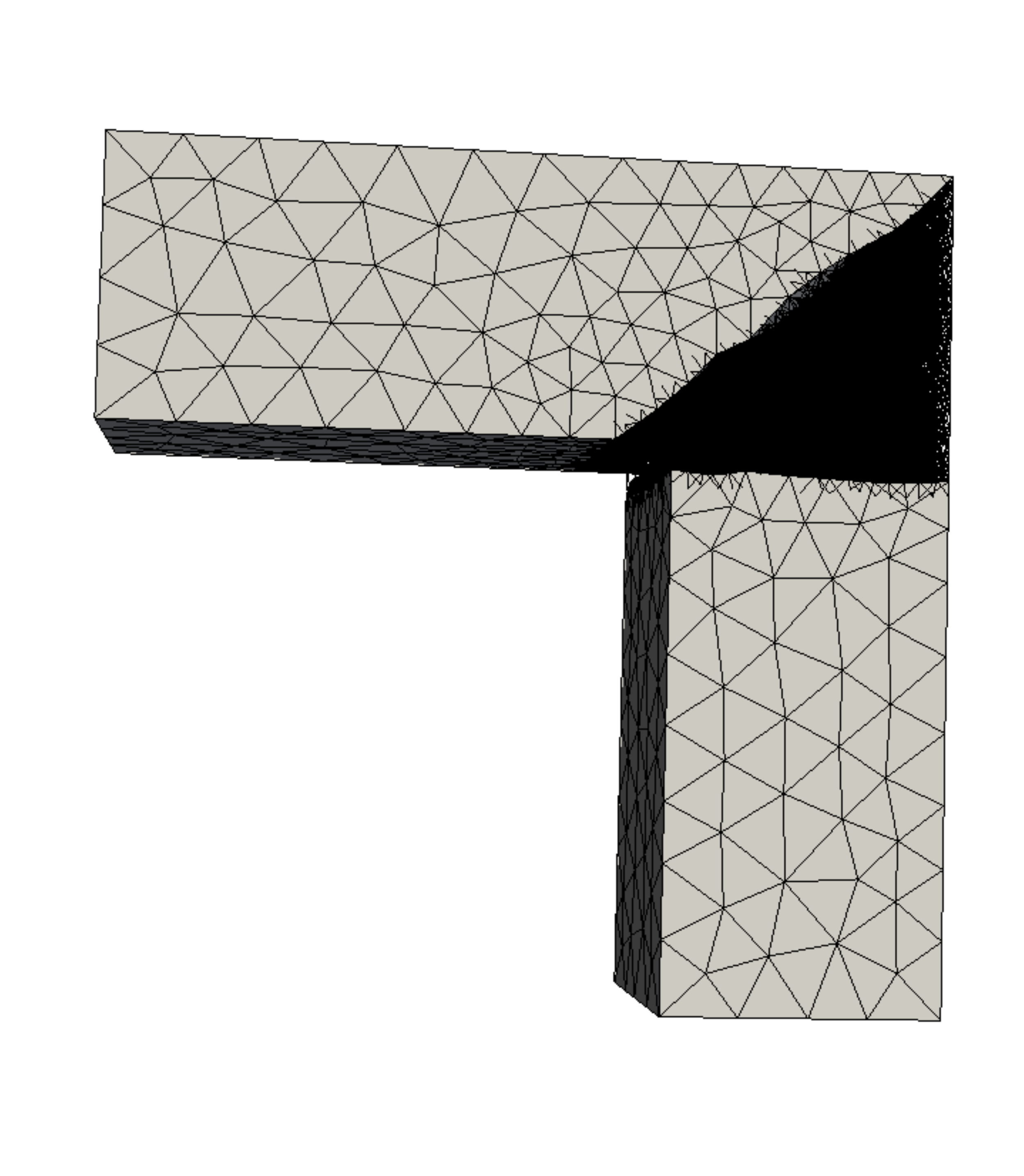}
    \caption{Geometry of interpolation mesh and the fully resolved domain.}
    \label{fig:3dCracking-geometry-2000}
  \end{center}
\end{figure}

\begin{figure}
  \begin{center}
    \includegraphics[trim=0cm 2.0cm 0cm 2.0cm, clip=true,width=1.0\textwidth]{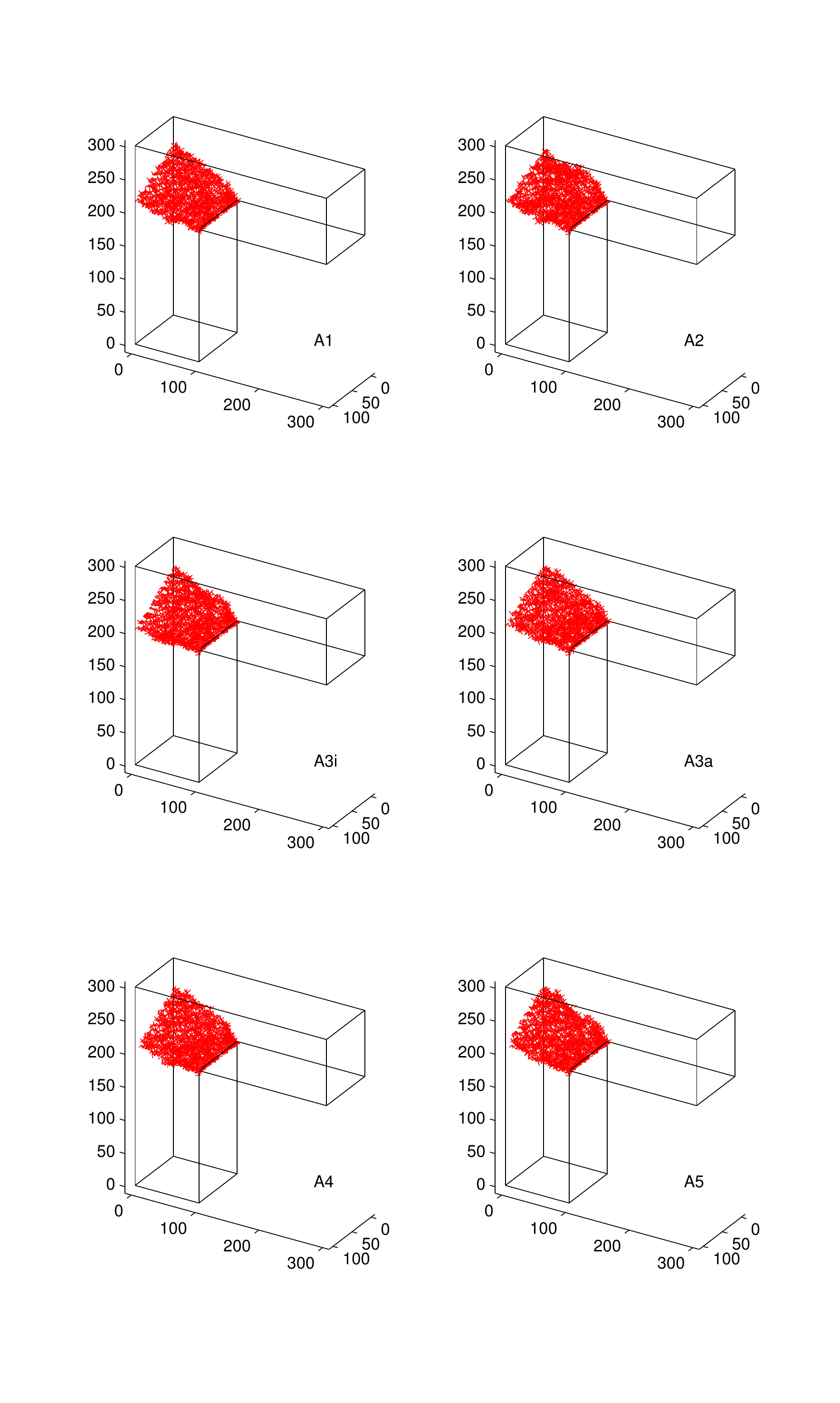}
    \caption{Cracked links computed with all approaches A1-A5. 
    }
    \label{fig:3dCracking-crack-33}
  \end{center}
\end{figure}

\begin{figure}
  \begin{center}
    \includegraphics[trim=0cm 2.0cm 0cm 2.0cm, clip=true,width=1.0\textwidth]{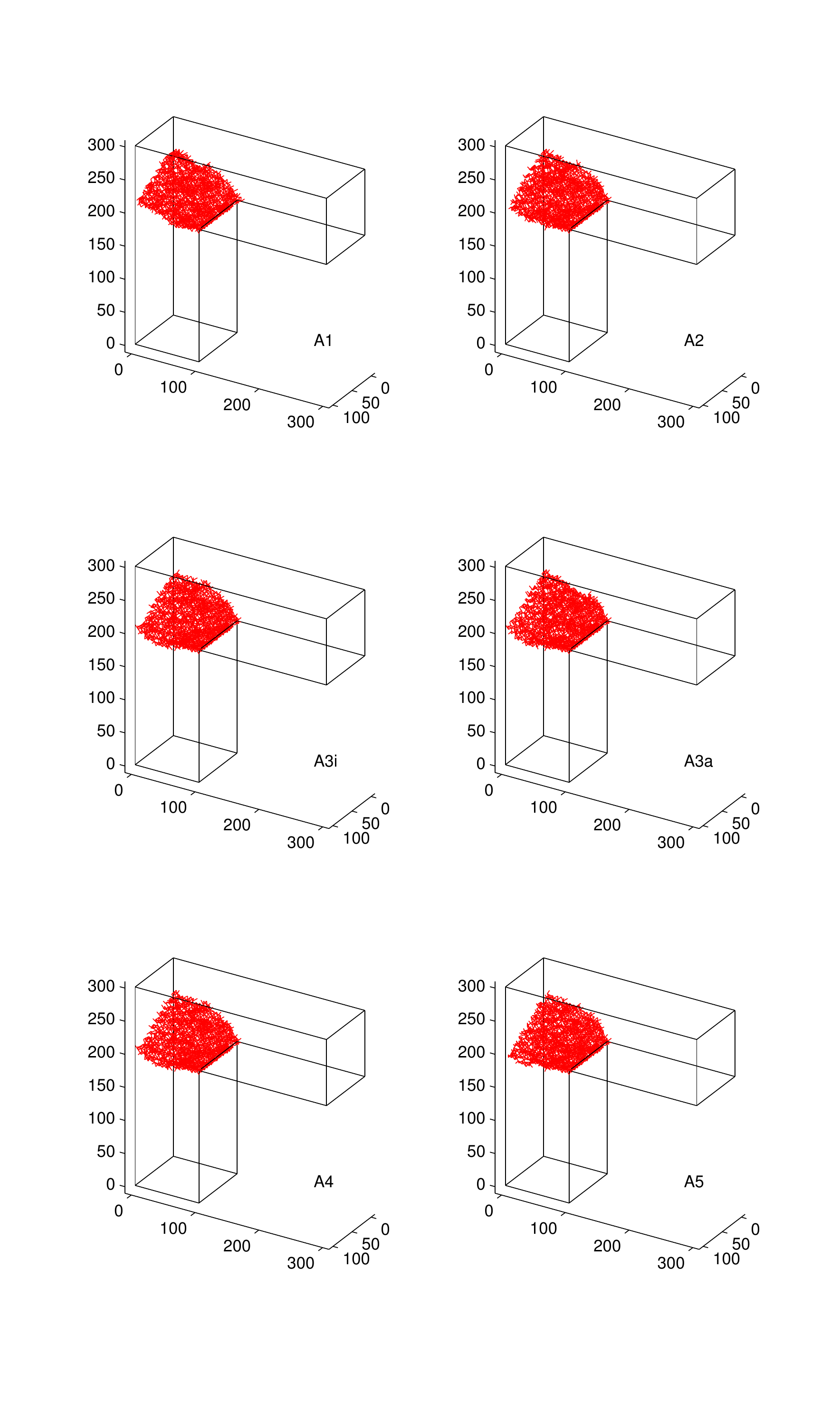}
    \caption{Cracked links computed with all approaches A1-A5. 
    }
    \label{fig:3dCracking-crack}
  \end{center}
\end{figure}

The cylindrical FSD used in the previous section is not convenient for simulations of a long crack trajectory.
Even if cracking of individual links outside the FSD is implemented, propagation of the crack outside the FSD is not accurate.
To be able to predict the correct crack trajectory, it is necessary to change the FSD adaptively or set up a priori an FSD in the region where crack propagation is expected.
For that purpose a new wedge-shaped FSD is selected; see Fig.~\ref{fig:3dCracking-geometry-2000}. For this FSD, cracking of 2000 links is computed and cracked links for approaches A1 and A5 are depicted in Figs.~\ref{fig:3dCracking-crack-33}--\ref{fig:3dCracking-crack}
for two random realizations of the underlying particle model. The list of cracked links for the A5 approach is not the same as for the exact approach A1, but the macroscopic shape of the crack is quite similar to the exact solution.

\begin{figure}
  \begin{center}
    \includegraphics[trim=0cm 0.0cm 0cm 0.0cm, clip=true,width=1.0\textwidth]{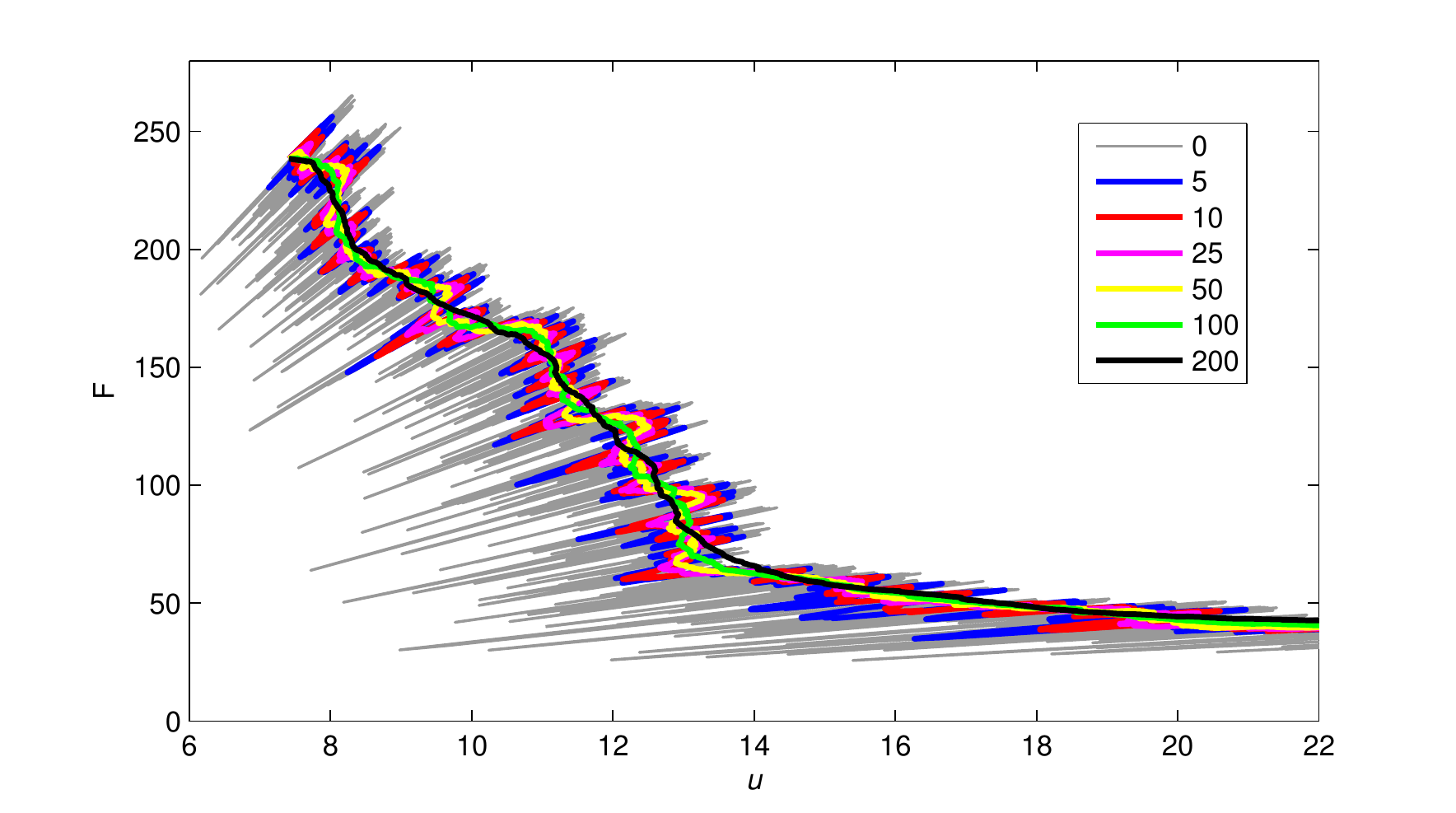}
    \caption{Descending branch of force-displacement diagram smoothed by averaging with different numbers of neighboring values, using a constant weight function.
    }
    \label{fig:FU-smoothing-const}
  \end{center}
\end{figure}

\begin{figure}
  \begin{center}
    \includegraphics[trim=0cm 0.0cm 0cm 0.0cm, clip=true,width=1.0\textwidth]{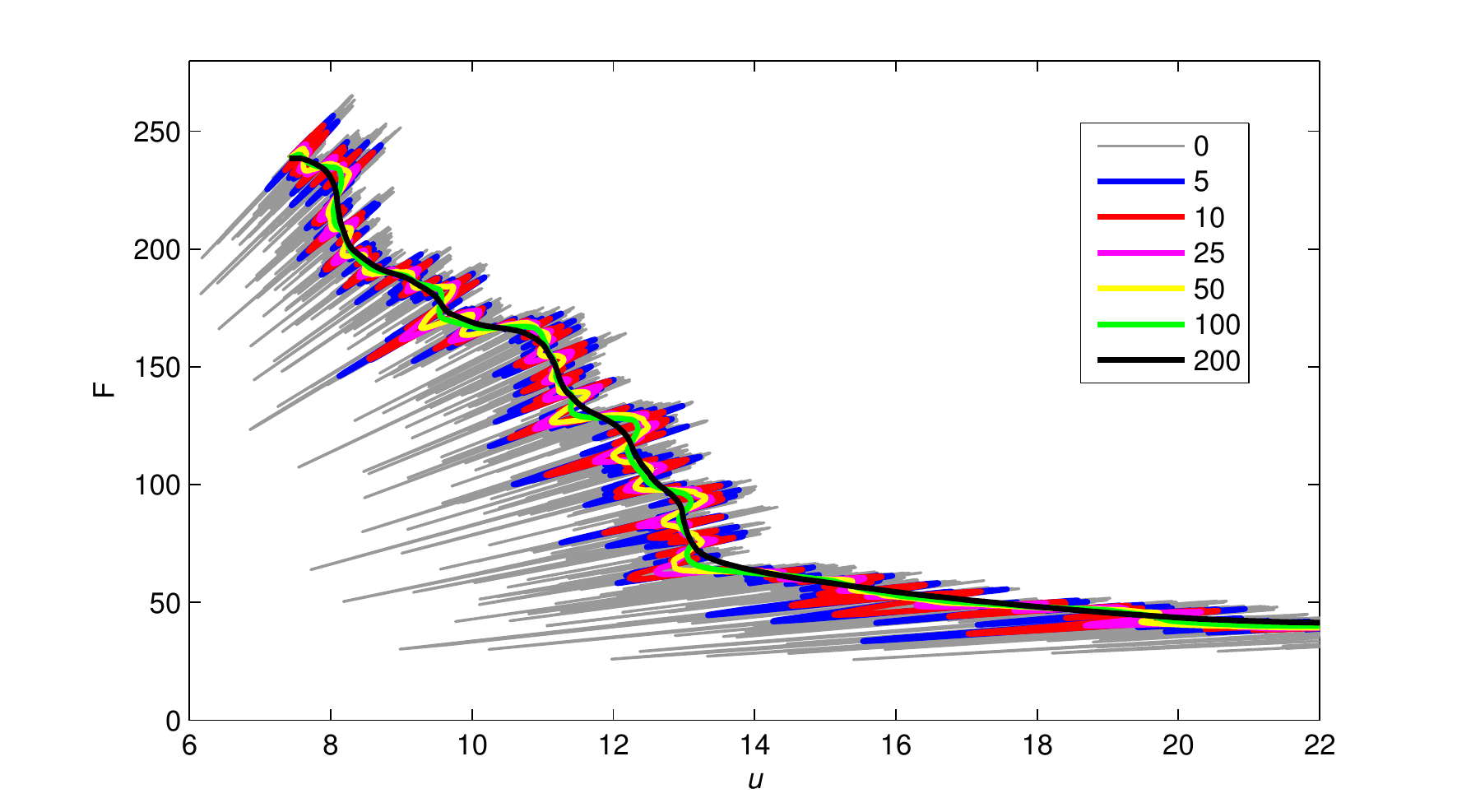}
    \caption{Descending branch of force-displacement diagram smoothed by averaging with different numbers of neighboring values, using a linear weight function.}
    \label{fig:FU-smoothing-lin}
  \end{center}
\end{figure}

The force-displacement diagram corresponding to a crack growth simulation in a microscopically elastic-brittle material exhibits high oscillations; see the scattered gray lines in Figs.~\ref{fig:FU-smoothing-const}--\ref{fig:FU-smoothing-lin}.
Interpretation and comparison of such force-displacement diagrams is facilitated
by smoothing the results.
The smoothing procedure replaces the values of force and displacement in each characteristic point by their weighted averages around this point. Different numbers of neighboring points have been used, ranging from $\pm$5 to $\pm$100. Three types of weight functions have been considered, namely constant, linear (closer points have a stronger influence) and bell-shaped polynomial (in the form often used by nonlocal material models) 
given by
\begin{equation}
	w(s) = \bigg \langle 1 - \frac{s^2}{R^2} \bigg \rangle^2
\end{equation}
where $s$ is one plus the number of points between the averaged point and the point for which the weight is evaluated. Parameter $R$ equals to one plus the number of considered neighboring points, and $\langle\ldots\rangle$ are Macauley brackets denoting the positive part.
The results for different numbers of neighboring points are compared for constant and linear weight functions in Fig.~\ref{fig:FU-smoothing-const} and \ref{fig:FU-smoothing-lin}, respectively.
The results obtained with the bell-shaped weight function are very similar to the results with linear weights.
Constant weights lead to sharper shapes of the final diagrams and, for the same number of used neighboring points, the oscillations are smaller in comparison with linear or polynomial weights.
All three variants of smoothing reflect the character of the original diagram and
reduce oscillations.

\begin{figure}
  \begin{center}
    \includegraphics[trim=0cm 0.0cm 0cm 0.0cm, clip=true,width=1.0\textwidth]{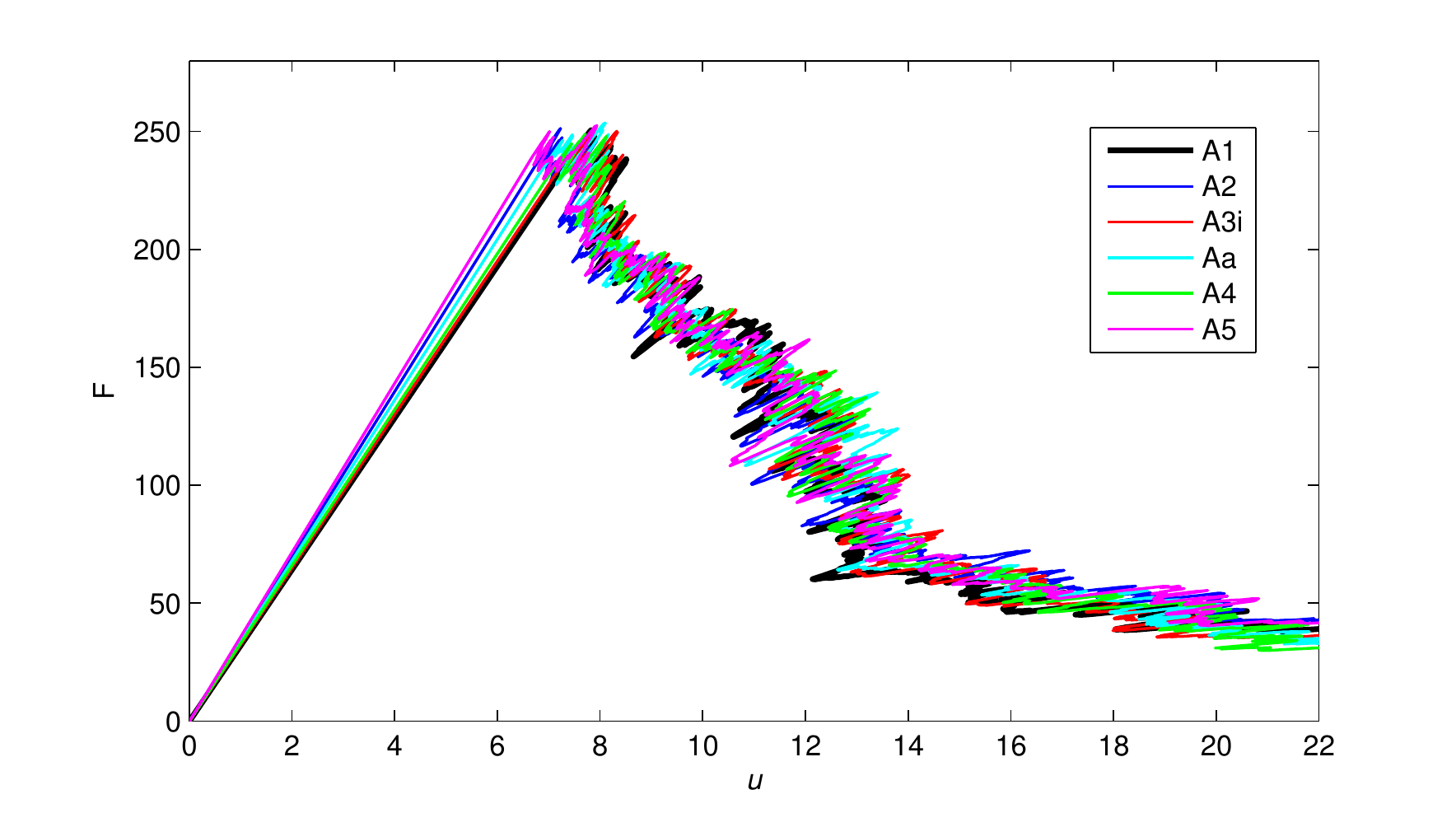}
    \caption{Force-displacement diagrams for microstructure with 3000 cracked links computed according to different approaches and smoothed by averaging $\pm$10 neighboring values with constant weight functions.}
    \label{fig:FU-const-10}
  \end{center}
\end{figure}

\begin{figure}
  \begin{center}
    \includegraphics[trim=0cm 0.0cm 0cm 0.0cm, clip=true,width=1.0\textwidth]{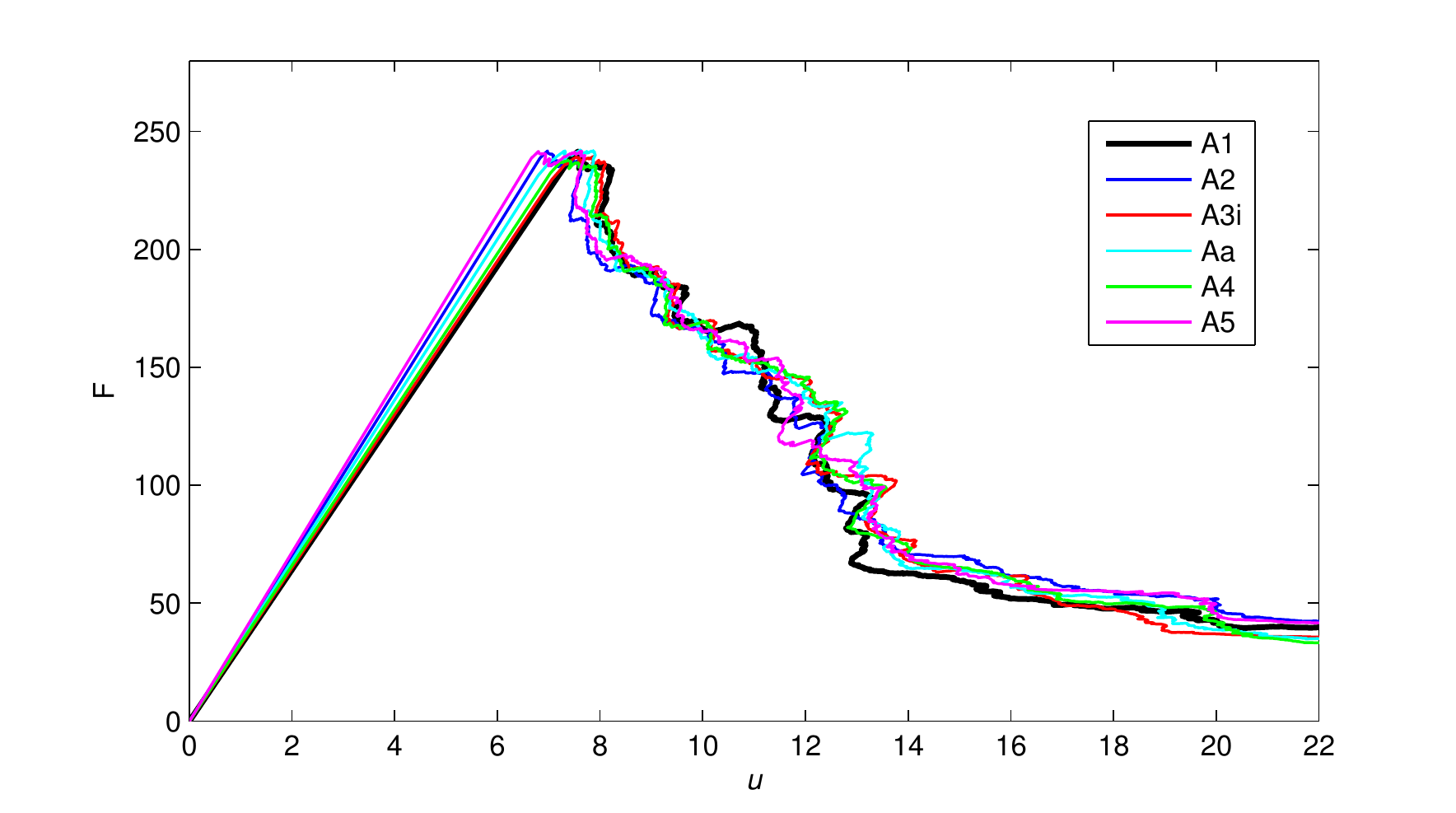}
    \caption{Force-displacement diagrams for microstructure with 3000 cracked links computed according to different approaches and smoothed by averaging $\pm$50 neighboring values, using a constant weight function.}
    \label{fig:FU-const-50}
  \end{center}
\end{figure}

\begin{figure}
  \begin{center}
    \includegraphics[trim=0cm 0.0cm 0cm 0.0cm, clip=true,width=1.0\textwidth]{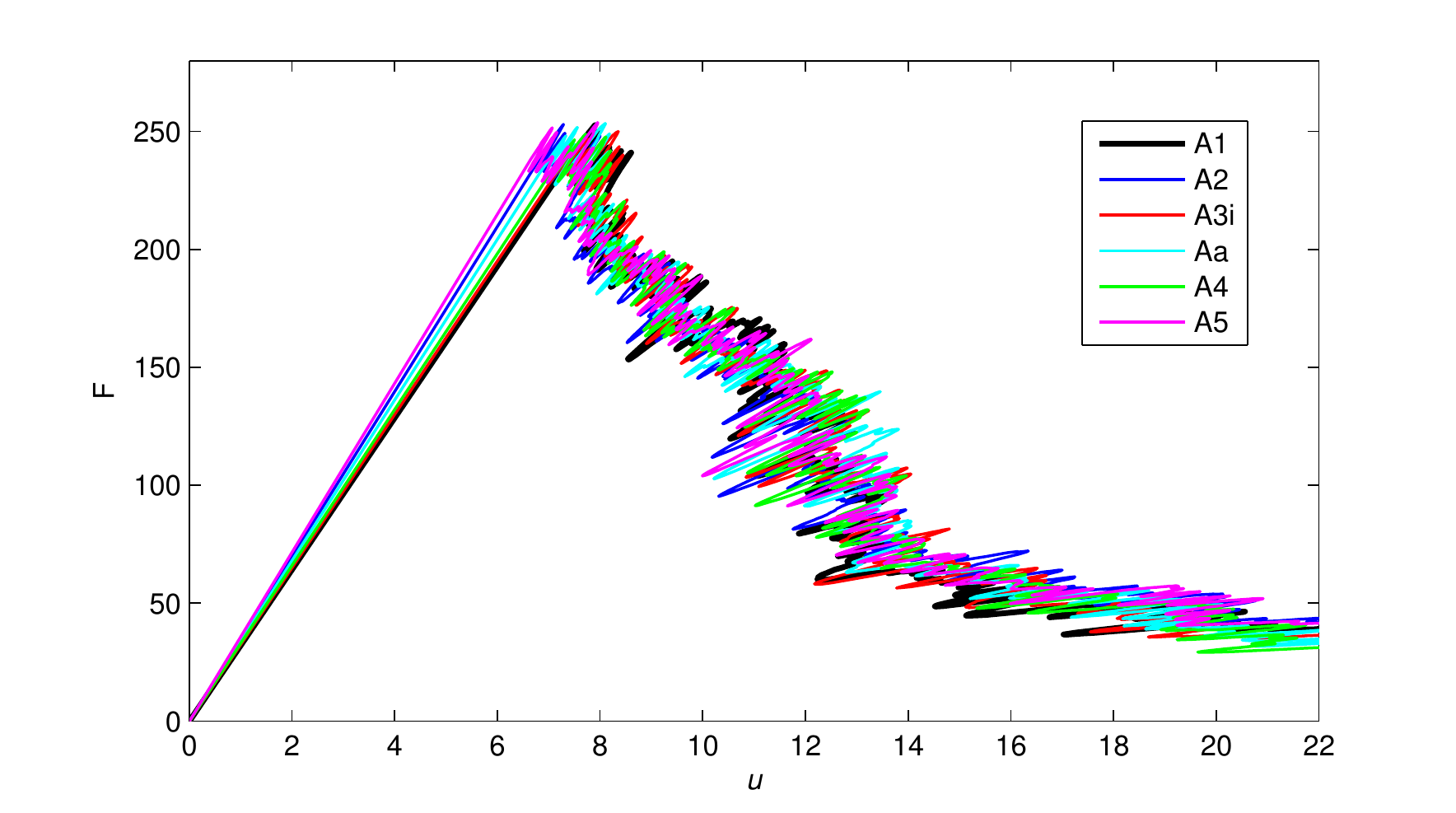}
    \caption{Force-displacement diagrams for microstructure with 3000 cracked links computed according to different approaches and smoothed by averaging $\pm$10 neighboring values, using a linear weight function.}
    \label{fig:FU-lin-10}
  \end{center}
\end{figure}

\begin{figure}
  \begin{center}
    \includegraphics[trim=0cm 0.0cm 0cm 0.0cm, clip=true,width=1.0\textwidth]{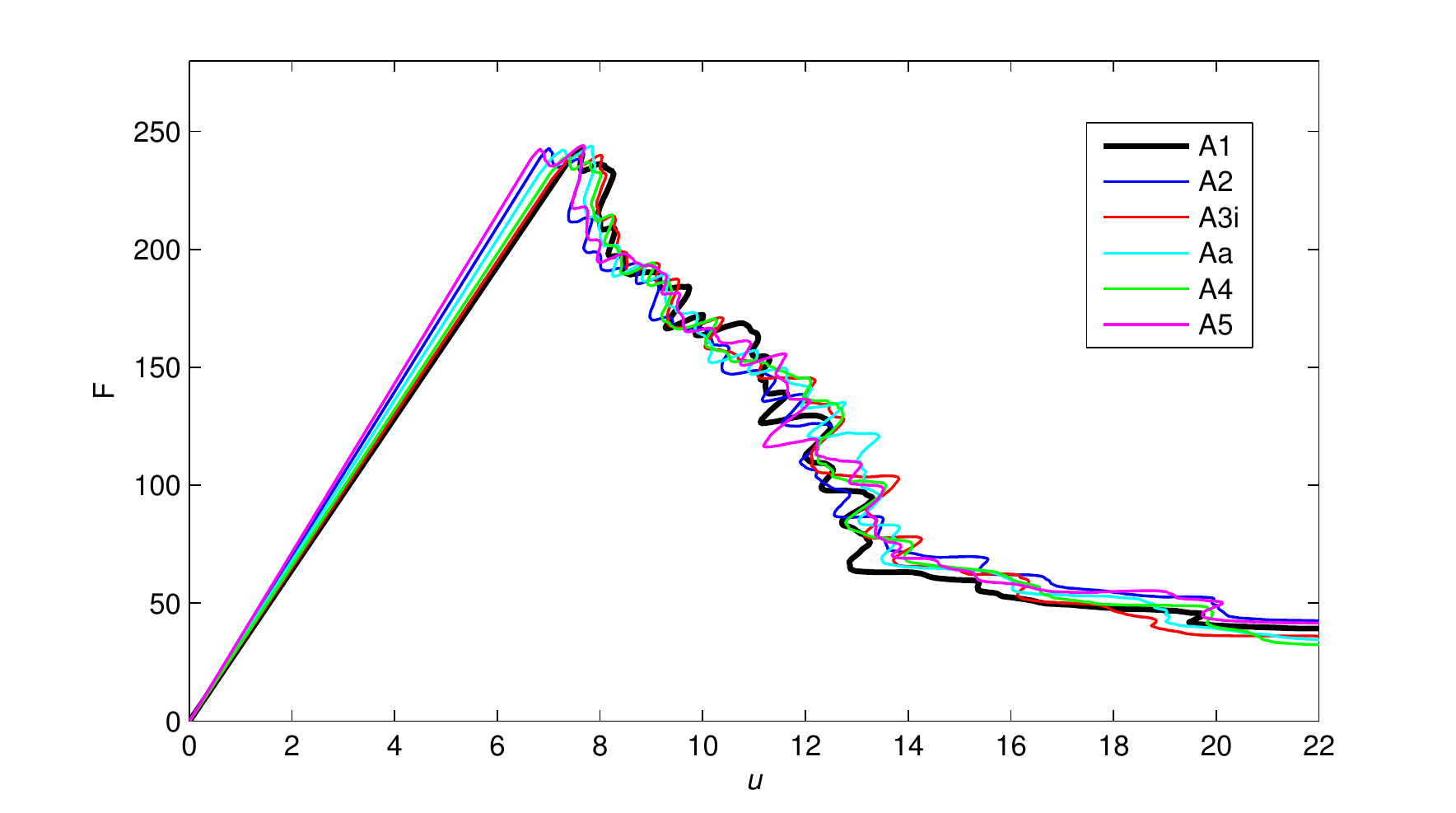}
    \caption{Force-displacement diagrams for microstructure with 3000 cracked links computed according to different approaches and smoothed by averaging $\pm$50 neighboring values, using a linear weight function.}
    \label{fig:FU-lin-50}
  \end{center}
\end{figure}

Diagrams for all approaches smoothed with constant weights are compared in Fig.~\ref{fig:FU-const-10} and \ref{fig:FU-const-50} for 10 and 50 neighboring points.
The same diagrams smoothed with linear weights are compared in Fig.~\ref{fig:FU-lin-10} and \ref{fig:FU-lin-50}.

In accordance with patch tests, the initial elastic response of the simplified approaches is stiffer than the exact solution. 
However, the shape of the softening branch is captured by all approaches very well.


\begin{table}						
\begin{center}							
\begin{tabular}{|l|r|r|r|}							
\hline							
~	&	\multicolumn{1}{|c|}{$F_{\rm max}$ error}	&	\multicolumn{1}{|c|}{${F_{\rm max}}$ displacement}	&	\multicolumn{1}{|c|}{$F_{\rm max}$ step}\\
~	&	\multicolumn{1}{|c|}{[\%]}	&	\multicolumn{1}{|c|}{error [\%]}	&	\multicolumn{1}{|c|}{number [\%]}\\ \hline
A1  	&	 -			&	 -			&	13 / 36\\
A2  	&	0.12 / 2.28		&	-7.91 / -6.77	&	11 / 15\\
A3i  	&	-1.30 / 2.15	&	-1.13 / -0.05	&	32 / 17\\
A3a 	&	-0.38 / 4.01	&	-4.61 / -2.65	&	32 / 18\\
A4  	&	-0.77 / 0.80	&	-2.17 / -2.73	&	32 / 17\\
A5  	&	0.18 / 3.31		&	-9.68 / -7.94	&	18 / 18\\ \hline
\end{tabular}							
\caption{Tensile strength errors for two different random microstructures.}							
\label{tab:3d-L-beam-v-maxF}							
\end{center}							
\end{table}	

\subsection{Large-scale computations}

The maximum size of the problem to be solved is limited by the size of available memory. Typically, today's office PCs come with 4GB of RAM. For such a computer, the maximum size of the L-shaped specimen that can be solved by OOFEM depends on the fineness of the microstructure. For a direct solver with a symmetric skyline matrix storage format, the maximum fineness of microstructure that can be solved
using the pure particle model (approach A1) is 27 particles along the shortest edge, which leads to 282,852 DOFs and 820,846 links. The more powerful conjugate gradient iterative solver with a dynamically growing compressed column matrix storage format is able to solve examples with 46 particles along the shortest edge and with 1,424,068 DOFs and 4,190,040 links. Larger examples cannot be run on a single standard PC due to lack of memory.

QC-based approaches can handle particle models with microstructure density up to 89 particles along the shortest edge using the same cheap PC with 4GB of RAM.
The full particle model has 3,493,161 particles, 31,007,592 links and 10,439,878 DOFs.
The QC approach uses a fully resolved domain with 433,265 repnodes and 3,694,490 links that cover 11.8 \% of the solved domain.
In the remaining part of the domain, an interpolation mesh consisting of 4,457 elements and 1,098 mesh nodes is used,
which results into a total of 3,698,947 elements and 1,302,939 DOFs.
In this simplified solution, the link with the maximum strain is predicted correctly and the relative error in maximum strain is just a few percent.

Modern supercomputers would allow to solve problems with 
a finer microstructure but such computations can be quite expensive. Furthermore, even for supercomputers, a finite limit on the size of the problem always exists and QC-based approaches can make even larger systems solvable.

\section{Conclusions and future work}

The presented example has demonstrated that QC-based methods can lead to a substantial reduction of the computational cost. The error induced by this reduction can be kept
within acceptable limits by suitably setting the region of high interest (fully resolved domain, FSD). 

Macroscopic properties associated with the global stiffness are naturally affected by a certain error induced by interpolation.
Local phenomena such as cracking are well captured by sufficiently large FSDs.

Approaches A3i and A4 based on isotropic homogenization tend to underestimate the global stiffness if the material is significantly anisotropic. In such cases, these approaches seemingly appear to be more accurate in certain examples but accuracy and convergence of this approach are not guaranteed.
On the other hand, the A5 approach based on local anisotropic homogenization seems to be very powerful. The homogenization error of A5 is negligible and this approach provides almost as accurate results as A2 while running substantially 
faster.

Both elastic and simple inelastic material models have been presented. So far, the inelastic behavior has been considered to have the form of brittle failure on the
microscopic level.
Future work will deal with optimization of efficiency and extensions to elastoplasticity and to softening material response, e.g., to damage-based models.

\section*{Acknowledgement}
Financial support received 
from the Czech Science Foundation (GA\v{C}R project No.~14-00420S) is gratefully acknowledged.


\begin{thebibliography}{99}

\topsep=0.0ex
\parsep=0.0ex
\parskip=0.0ex
\itemsep=0.0ex



\bibitem{Liu10} J. Liu, Z. Chen, and K. Li,
``A 2-d lattice model for simulating the failure of paper'',
Theoretical and Applied Fracture Mechanics, 54(1), 1-10, 2010.

\bibitem{BeexVer13} L. Beex, C. Verberne, and R. Peerlings,
``Experimental identification of a lattice model for woven fabrics: Application to electronic textile'',
Composites Part A: Applied Science and Manufacturing, 48, 82-92, 2013.

\bibitem{BeePee15} L. Beex, R. Peerlings, K. van Os and M. Geers,
``The mechanical reliability of an electronic textile investigated using the virtual-power-based quasicontinuum method'',
Mechanics of Materials, 80, 52–66, 2015.

\bibitem{WilBeex13} D. Wilbrink, L. Beex, and R. Peerlings,
``A discrete network model for bond failure and frictional sliding in fibrous materials'',
International Journal of Solids and Structures, 50(9), 1354-1363, 2013.

\bibitem{RidGon10} A. Ridruejo, C. Gonz\'{a}lez and J. LLorca,
``Damage micromechanisms and notch sensitivity of glass-fiber non-woven felts: An experimental and numerical study'',
Journal of the Mechanics and Physics of Solids, 58, 1628-1645, 2010.

\bibitem{KulaUesa12} A. Kulachenko and T. Uesaka,
``Direct simulations of fiber network deformation and failure''
Mechanics of Materials, 51, 1-14, 2012.

\bibitem{PengCao04} X.Q. Peng and J. Cao,
``A continuum mechanics-based non-orthogonal constitutive model for woven composite fabrics'',
Composites: Part A: Applied science and manufacturing, 36, 859-874, 2005.

\bibitem{Baz90} Z. P. Ba\v{z}ant, M. R. Tabbara, M. T. Kazemi, and G. Pijaudier-Cabot,
``Random particle model for fracture of aggregate or fiber composites'',
Journal of Engineering Mechanics, 116(8), 1686–1705, 1990.

\bibitem{Jin16} C. Jin, N. Buratti, M. Stacchini, M. Savoia and G. Cusatis,
``Lattice discrete particle modeling of fiber reinforced concrete: Experiments and simulations'',
European Journal of Mechanics-A/Solids, 57, 85-107, 2016.

\bibitem{CusPel11} G. Cusatis, D. Pelessone, and A. Mencarelli,
``Lattice discrete particle model (ldpm) for failure behavior of concrete i: Theory'',
Cement and Concrete Composites, 33(9), 881-890, 2011.

\bibitem{LilMie03} G. Lilliu and J. van Mier,
``3d lattice type fracture model for concrete'',
Engineering Fracture Mechanics, 70(7-8), 927-941, 2003.

\bibitem{LiuDen07} J. Liu, S. Deng, J. Zhang, and N. Liang. Lattice type of fracture model for concrete. Theoretical and Applied Fracture Mechanics, 48(3):269 – 284, 2007.

\bibitem{Baz10} Z. P. Ba\v{z}ant.
``Can multiscale-multiphysics methods predict softening damage and structural failure?'',
International Journal for Multiscale Computational Engineering, 8(1), 2010.

\bibitem{CurMil03} W. A. Curtin and R. E. Miller,
``Atomistic/continuum coupling in computational materials science'',
Modelling and Simulation in Materials Science and Engineering, 11, R33-R68, 2003.

\bibitem{TadPhi96} E. Tadmor, R. Phillips, and M. Ortiz,
``Mixed atomistic and continuum models of deformation in solids'',
Langmuir, 12, 4529-4534, 1996.

\bibitem{TadOrt96} E. B. Tadmor, M. Ortiz, and R. Phillips,
``Quasicontinuum analysis of defects in solids'',
Philosophical Magazine A, 73, 1529-1563, 1996.

\bibitem{TadMil05} E. Tadmor and R. Miller,
``The theory and implementation of the quasicontinuum method'',
Handbook of Materials Modeling, Springer Netherlands, 663-682, 2005.

\bibitem{BeePee14} L. Beex, R. Peerlings, and M. Geers,
``Central summation in the quasicontinuum method'',
Journal of the Mechanics and Physics of Solids, 70, 242-261, 2014.

\bibitem{BeePee14a} L. Beex, R. Peerlings, and M. Geers,
``A multiscale quasicontinuum method for dissipative lattice models and discrete networks'',
Journal of the Mechanics and Physics of Solids, 64, 154-169, 2014.

\bibitem{BeePee2014c} L. Beex, R. Peerlings, and M. Geers,
``A multiscale quasicontinuum method for lattice models with bond failure and fiber sliding'',
Computer Methods in Applied Mechanics and Engineering, 269, 108-122, 2014.

\bibitem{beex15} L. A. A. Beex, O. Roko\v{s}, J. Zeman and S. P. A. Bordas,
``Higher‐order quasicontinuum methods for elastic and dissipative lattice models: uniaxial deformation and pure bending'',
GAMM‐Mitteilungen, 38(2), 344-368, 2015.

\bibitem{MilTad09} R. E. Miller and E. B. Tadmor,
``A unified framework and performance benchmark of fourteen multiscale atomistic/continuum coupling methods'',
Modelling and Simulation in Materials Science and Engineering, 17, 053001, 2009.

\bibitem{MilTad02} R. E. Miller and E. B. Tadmor, 
``The quasicontinuum method: Overview, applications and current directions'', 
Journal of Computer-Aided Materials Design, 9, 203-239, 2002.

\bibitem{TadMil11} E. B. Tadmor and R. E. Miller,
``Modeling Materials: Continuum, Atomistic and Multiscale Techniques'',
Cambridge University Press, 2011.

\bibitem{rokos16} O. Roko\v{s},  L. A. Beex, J. Zeman and R. H. Peerlings,
``A Variational Formulation of Dissipative Quasicontinuum Methods'',
in press... arXiv preprint arXiv:1601.00625, 2016.


\bibitem{MamLar15} A. Memarnahavandi, F. Larsson and K. Runesson,
``A goal-oriented adaptive procedure for the quasi-continuum method with cluster approximation'',
Computational Mechanics, 55, 617-642, 2015.

\bibitem{koch14} D. M. Kochmann and G. N. Venturini,
``A meshless quasicontinuum method based on local maximum-entropy interpolation'',
Modelling and Simulation in Materials Science and Engineering, 22, 2014.

\bibitem{MikJir15} K. Mike\v{s} and M. Jir\'{a}sek,
The quasicontinuum method extended to disordered materials,
in Proceedings of the 15th International Conference on Civil, Structural and
Environmental Engineering Computing, Civil-Comp Press, Stirlingshire, Scotland, 2015.

\bibitem{oofem01} B. Patz\'{a}k and Z. Bittnar
``Design of object oriented finite element code.''
Advances in Engineering Software, 32(10), 759-767, 2001.

\bibitem{oofem12} B. Patz\'{a}k and D. Rypl
``Object-oriented, parallel finite element framework with dynamic load balancing.''
Advances in Engineering Software, 47(1), 35-50, 2012.

\bibitem{Pat12} B. Patz\'{a}k, 
``OOFEM -- an object-oriented simulation tool for advanced modeling of materials and structures'', 
Acta Polytechnica, 52, 59-66, 2012.

\bibitem{SheMil99} V. Shenoy, R. E. Miller, E. B. Tadmor, D. Rodney, R. Phillips, and M. Ortiz,
``An adaptive finite element approach to atomic-scale mechanics—the quasicontinuum method'',
Journal of the Mechanics and Physics of Solids, 47, 611-642, 1999.

\bibitem{Ryp15} D. Rypl,
T3D mesh generator,
[online]. [cit. 2015-5-1]. Available at: http://ksm.fsv.cvut.cz/dr/t3d.html.



\bibitem{Hill63} R. Hill,
``Elastic properties of reinforced solids: some theoretical principles''
Journal of the Mechanics and Physics of Solids 11(5), 357-372, 1963.

\bibitem{Man72} J. Mandel,
``Plasticit\'{e} classique et viscoplasticit\'{e}. International Centre for Mechanical Sciences, Courses and Lectures No. 97.'' Springer, Udine 1972.


\bibitem{SteiEliz06} P. Steinmann, A. Elizondo and R. Sunyk,
``Studies of validity of the Cauchy-Born rule by direct comparison of continuum and atomistic modelling. Modelling and Simulation in Materials'', Science and Engineering, 15(1), S271, 2006.

\bibitem{Ozbolt01} J. O\v{z}bolt, Y. Li, and I. Ko\v{z}ar,
``Microplane model for concrete with relaxed kinematic constraint'', International Journal of Solids and Structures, 38(16), 2683-2711, 2001.


\end{thebibliography}
\end{document}